\documentclass[11pt]{article}

\usepackage{graphicx}

\usepackage[super,numbers,sort&compress]{natbib}

\usepackage{epsfig}
\usepackage{cite}
\usepackage{pifont}
\usepackage{amssymb}
\usepackage{latexsym}
\usepackage{verbatim}
\usepackage{array}
\usepackage{xspace}
\usepackage{colordvi}
\usepackage{wasysym}
\usepackage{marvosym}
\usepackage{setspace}
\usepackage{hhline}

\newlength{\saveparindent}
\newcommand{\rl}[1]{#1_{12}}
\newcommand{\nn}{\\ \nonumber}
\newcommand{\rhat}{\hat{\textbf{r}}}

\newcounter{saveeqn}%
\newcommand{\alpheqn}{\setcounter{saveeqn}{\value{equation}}%
\stepcounter{saveeqn}\setcounter{equation}{0}%
\renewcommand{\theequation}{\mbox{\arabic{saveeqn}\alph{equation}}}}%
\newcommand{\reseteqn}{\setcounter{equation}{\value{saveeqn}}%
\renewcommand{\theequation}{\arabic{equation}}}%

\newcommand {\be}{\begin{equation}} 
\newcommand {\ee}{\end{equation}}

\def \be{\begin{equation}}
\def \ee{\end{equation}}
\def \bea{\begin{eqnarray}}
\def \eea{\end{eqnarray}}

\def \rA{{\bf r}_{\mbox{\tiny{A}}}}
\def \rB{{\bf r}_{\mbox{\tiny{B}}}}
\def \rN{r_{\mbox{\tiny{N}}}}
\def \rAN{{\bf r}_{\mbox{\tiny{A}}}^{\mbox{\tiny{N}}}}
\def \rBN{{\bf r}_{\mbox{\tiny{B}}}^{\mbox{\tiny{N}}}}

\def \lp{\ell_{\mbox{\tiny P}}}
\def \lD{\ell_{\mbox{\tiny D}}}

\def \kB{k_{\mbox{\tiny B}}}

\def \s{\sigma}

\def \eps{\epsilon}

\def \Veff{V_{\mbox{\scriptsize{eff}}}}

\usepackage{flafter} 
\usepackage[hypertex]{hyperref}[2003/11/30]
\usepackage{color}

\def\degrees{^{\circ}}
\definecolor{darkgreen}{RGB}{0, 100, 0}

\definecolor{darkbrown}{RGB}{139, 69, 19}
\definecolor{teal}{RGB}{0, 128, 128}

\definecolor{purple}{RGB}{135,31,120}

\definecolor{orange}{RGB}{255,69,0}

\newcommand{\epssub}[1]{\epsilon_{\tiny{\mbox{#1}}}}

\newcommand{\boldgreek}[1]{\mbox{\boldmath$#1$}}

\def\thesection       {\Roman{section}}

\usepackage{enumerate}

\begin{document}

\bibliographystyle{aip}

\title{A systematically coarse-grained model for DNA, and its predictions for
persistence length, stacking, twist, and chirality}

\author{Alex~Morriss-Andrews, Joerg Rottler and
	 Steven~S.~Plotkin$^\ast$ \\Department of Physics and
Astronomy, University of British Columbia\\ 6224 Agricultural
Road, Vancouver, BC V6T1Z1, Canada. \\ \\
$^\ast$steve@phas.ubc.ca}

\date{}

\pagestyle{myheadings}
\markright{Coarse-graining DNA}

\pagenumbering{arabic}
\setcounter{page}{1}

\maketitle
\normalsize

\abstract{We introduce a coarse-grained model of DNA with bases
modeled as rigid-body ellipsoids to capture their anisotropic
stereochemistry. Interaction potentials are all physicochemical
and generated from all-atom simulation/parameterization with
minimal phenomenology. Persistence length, degree of stacking,
and twist are studied by molecular dynamics simulation as
functions of temperature, salt concentration, sequence,
interaction potential strength, and local position along the
chain, for both single- and double-stranded DNA where
appropriate. The model of DNA shows several phase transitions
and crossover regimes in addition to dehybridization, including
unstacking, untwisting, and collapse which affect mechanical
properties such as rigidity and persistence length. The model
also exhibits chirality with a stable right-handed and
metastable left-handed helix.}

\newpage
\section{INTRODUCTION}

DNA is likely the most well-studied biomolecule, with
structural, energetic, and kinetic characterization spanning
over 6 decades of research since its isolation as the carrier
of genetic information by Oswald Avery and
co-workers~\citep{AveryO44}. Elucidating the varied behaviors
of DNA has been primarily an experimental endeavor, due in
large part to the difficulties in capturing the molecule's
complex motion and function either computationally or
theoretically. The computational difficulties are primarily due
to the fact that much of the interesting behavior takes place
on time-scales 3-6 orders of magnitude longer than the longest
all-atom simulations of a system comparable to the typical size
of a gene (currently nanoseconds~\citep{CheathamTE04}).

For example, RNA polymerase transcribes DNA at a rate of about
$14 \: \mbox{ms/nucleotide}$ in Eukaryotes~\citep{DarzacqX07}
during elongation, with comparable rates in E.
Coli~\citep{YinH95}. These rates are further slowed by
transcriptional pausing to regulate arrest and
termination~\citep{GreiveSJ05,DavenportRJ00,HerbertKM06} by
$\sim$seconds per pause. Even ``fast'' processes such as
bacteriophage DNA ejection have translocation times greater
than $10\mu\mbox{s}/\mbox{bp}$~\citep{MangenotS05,GraysonP07}.
Time-scales for DNA packaging into the viral capsid are $\sim10
\mbox{ms}/\mbox{bp}$~\citep{SmithDE01}. Nucleosome condensation
time-scales at {\it in vivo} histone concentrations are $\sim
10 \mbox{ms}$~\citep{WagnerG05}. To address any of these
biologically relevant phenomena computationally currently
requires the introduction of coarse-grained models.  The choice
of coarse-grained DNA model reflects the empirical phenomena
the model intends to capture.  For example, a description of
Zinc-finger protein binding to DNA would require an accurate
representation of major and minor grooves, whereas a
description of the sequence-dependence of nanopore
translocation would require an accurate description of the
stereochemistry of bases. As in the above examples, these
phenomena also exhibit slow kinetics compared to time-scales
accessible computationally.  Binding rates for transcription
factors~\citep{NalefskiEA06} are on the order of $1/ms$ per
promoter at a cellular concentration of $\sim
10^6/\mbox{nucleus}$, and translocation times for
single-stranded DNA at typical experimental voltages are on the
order of $100 \mu s$ for a $100$bp sequence~\citep{MellerA00}.

On the coarsest scale, a piece of double-stranded DNA may be
described as a semiflexible polymer using the wormlike chain
model. On this level, the only parameters characterizing the
molecule are its total length and its bending rigidity
$\kappa$, which determines the persistence length
$\ell_p=\kappa/\kB T$. This model is very successful in
capturing chain properties on scales larger than $\ell_p$, for
instance the force-extension relation~\citep{MarkoJF95}, but
carries no information about the internal structure of the
double strand.  Slightly more refined models approximate
single-stranded DNA as a semiflexible chain of sterically
repulsive spheres~\citep{MatysiakS06,LuoK07}, which may carry
charge and thus interact with an external field, although
Coulombic monomer-monomer interactions were neglected in these
models. One bead per nucleotide representations of DNA have
also been used to describe supercoiling and local denaturation
in plasmids~\citep{TrovatoF08}. Bead-spring models of
double-stranded DNA with nonlinear interchain coupling through
hydrogen bonds have been used to study vibrational energy
transport and localization~\citep{TecheraM89}. An extension of
the bead-spring idea, where base-pairs were represented by a
planar collection of $14$ harmonically-coupled beads, allowed
for an investigation of spontaneous helix formation from
initial ladder-like conformations~\citep{TepperHL05}.  More
detailed representations have been proposed that describe DNA
on the level of spherically-symmetric base monomers attached to
a chain of similar monomers representing the
backbone~\citep{ZhangF95,DrukkerK01,Knotts07}.  In these
approaches base molecules and repeat units on the backbone are
modeled as spheres that interact with other bases through
phenomenological potentials mimicking van der Waals and
hydrogen bonds. On this level of resolution, it is for example
possible to study thermal denaturation of DNA, as well as
mechanical properties such as bending rigidity. Zhang and
Collins~\citep{ZhangF95} treated base-ribose moieties as a
single rigid body with residues at the positions of hydrogen
bonding heavy atoms. While the rigid bodies were constrained to
undergo two-dimensional motion, base-pairing through a modified
Morse potential allowed for a systematic study of the
structural changes during thermal denaturation. More recently,
Knotts {\it et al.}~\citep{Knotts07} introduced a refined
bead-spring model that distinguishes between sugar and
phosphate groups on the backbone and individual base molecules.
After parameter adjustment which included introducing a
G\={o}-like potential~\citep{Ueda78} to bias the system to the
crystal structure, the model was able to describe several
features of DNA physics such as the dependence of $\ell_p$ and
duplex stability on ionic concentration, as well as the
dynamics of thermally activated bubble formation in
hybridization.

For phenomena such as protein-DNA binding and DNA translocation, the
base stereochemistry characterizing the DNA sequence is of particular
importance.  To this end, we introduce a model of DNA where phosphate
and sugar groups are represented by one coarse-grained (spherical)
residue, and bases by a rigid-body ellipsoid.  Sterically, bases of
DNA more accurately resemble a flat plate than a spherical object.
Energetically, base stacking interactions, predominantly governed by
electron correlation (van der Waals) interactions~\citep{LuoR01}, play
a significant if not dominant role in the stability of the double
helix~\citep{YakovchukP06}.  We adopt a systematic coarse-graining
approach, where we parameterize the effective interactions through
all-atom simulations wherever possible. The interactions describing
the stacking of bases are optimized against a fully atomistic
representation of the base residues. Interactions describing covalent
bonds, angles and dihedrals along the DNA strand are obtained from
equilibrium simulations of a short all-atom strand of DNA. One central
difference in the present model from previous models is therefore the
absence of any structure-based potentials, i.e.  we do not use
effective G\={o}-like potentials to bias the system toward an
experimentally determined structure. All behavior in the present model
is a direct consequence of physicochemical interactions and the
systematic parameterization procedure.

The definition of the model and details of our coarse-graining scheme
are described in the subsequent Model section and in the Supplementary
Material (SM can be found at \url{http://www.physics.ubc.ca/~steve/publication/23-suppmat.pdf}). We then investigate the predictions of the model for
several important properties of DNA, in particular persistence length
$\ell_p$ and radius of gyration $R_g$ of both double and single stranded
DNA. We introduce intuitive and quantitative methods to calculate both
helical twist and the stacking of bases as order parameters
characterizing the degree of native DNA structure.  The behavior of
these quantities is quantified as various internal and environmental
factors are varied, including sequence, interaction strength, strand
length, temperature, and ionic concentration.

\section{MODEL}

\begin{figure}[!h]
\begin{center}
\begin{tabular}{cc}
      \includegraphics*[width=8cm]{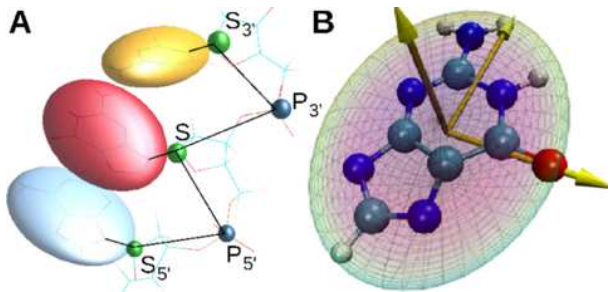}
\end{tabular}
      \caption[]{Our coarse-graining scheme overlaid onto an all atom
        visualization.  Panel (A) shows a section of 3 nucleotides:
        spherical sugar (S), and phosphate (P) residues, together with
        ellipsoidal bases. Panel (B) shows a close up of a base with
        ellipsoid overlay. The axes indicate the principal axes of the
        base, and are aligned with the principal axes of the
        ellipsoid, which determine the base's initial orientation.}
      \label{figdna}
\end{center}
\end{figure}

\subsection{Parameterization of the model}
Fig.~\ref{figdna} depicts a piece of DNA in our coarse-grained model
superimposed on the all-atom representation. Alternating sugar (S) and
phosphate (P) groups are replaced by spherically-symmetric
residues. The sugar residue is located at the position of the C1$'$
atom in standard PDB notation, which is connected to the base by a
single covalent bond. The phosphate residue is located at the center
of mass of the phosphate group ($PO_4$).  The bases Adenine, Cytosine,
Guanine, and Thymine are represented as ellipsoidal objects whose
structure and interaction potential are described below.

An immediate problem in coarse-grained systems is the loss of
knowledge regarding the Hamiltonian or energy function governing
dynamics in the system.  In all-atom classical molecular dynamics
simulations, this problem is resolved by characterizing quantities
such as partial charges and interaction potentials with empirical
parameters derived from best fit to combinations of experimental and
{\it ab initio} quantum mechanical target data~\citep{FoloppeN00}. We
wish to adopt a similar methodology with respect to the extraction of
phenomenological parameters here, by characterizing our coarse-grained
system in terms of effective parameters derived from all-atom
simulations.
The potential energy function for our system has the following form:
\bea
V &=& V_{RE^2} ({\bf B_1}, {\bf B_2}) + V_{RE^2} ({\bf B_1},\mbox{res}_2) +
V_{HB}({\bf B_1}, {\bf B_2}) + V_{LJ}(r_{ss},r_{sp}) +  V_{C}(r_{pp})
\nonumber \\
& &+  V_{bond}(r) + V_{angle}(\theta) + V_{dihedral}(\phi)
\label{eqV}
\eea
The individual terms in Eq.~(\ref{eqV}) are
described below.

\subsection{Base-base van der Waals interactions}
The base-base interaction potentials $V_{RE^2} ({\bf B_1}, {\bf
B_2})$ must account for the shape and relative orientation of
the anisotropic ellipsoids representing the bases. We adopt the
functional form derived by Babadi {\it et
al}~\citep{BabadiM06}, which is a modification of Gay-Berne
potentials~\citep{GayJG81} to better capture the long-distance
convergence to all-atom force fields. This so-called RE$^2$
potential has somewhat complicated form and is summarized in
the Appendix \ref{secappendix}. Describing the effective
ellipsoids characterizing the bases involves several
geometrical and energetic parameters. These are the three
``half-diameters'' for each base along principal axes
($\s_x,\s_y,\s_z$), three corresponding ``reciprocal
well-depths'' for each base $(\eps_x,\eps_y,\eps_z)$, a
parameter $\s_c$ characterizing the interaction range between
atoms in the all-atom potential, and the Hamaker constant
$A_{12} = 4 \pi \epssub{LJ} \rho^2 \s^6$ in Lennard-Jones (LJ)
units (given by Eq.~(\ref{eqLJ})), where $\rho$ is the number
density or reciprocal of the effective volume of each
ellipsoid.  The above $14$ parameters are found for the $10$
possible base-base interactions (A-A, A-C, etc...) by best fit
between the RE$^2$ functional form and the all-atom MM3 force
field~\citep{LiiJH89} with Buckingham exponential-6 potentials,
following the parameterization procedure in~\citep{BabadiM06}.
The resulting values are used in $V_{RE^2}({\bf B_1}, {\bf
B_2})$ in Eq.~(\ref{eqV}), and are summarized in
Table~\ref*{S1-tabRE2} in SM.

Though the values in Table~\ref*{S1-tabRE2} are essential to
describe the potential, it is more useful intuitively to
determine the half-diameters of the bases by taking two
identical bases and aligning them along the three principal
axes. Two identical bases are brought together along a
principal axis until their $RE^2$ potential is equal to $1\kB
T$ and $10 \kB T$. The distances that result are measures of
the energetically-determined effective diameters of the
ellipsoid representing the base, and are tabulated for all
residues in Tables~\ref*{S1-tabsigmalike}
and~\ref*{S1-tabsigmaunlike} in SM.  The distances determined
in this manner from the $RE^2$ equipotentials should correlate
with the size of the bases as determined other independent
measures such as the effective hydrodynamic radii of the bases,
which can be extracted from all-atom simulation studies of the
diffusion of a base in water. This comparison is described
below and in the Appendix.

\subsection{Base-Sugar and Base-Phosphate van der Waals interactions}
Bases may interact as well with phosphate or sugar residues according
to so-called sphere-asphere potentials $V_{RE^2} ({\bf B_1},
\mbox{res}_2)$, which are a limiting case of $RE^2$ interactions when
one entity is spherical. The form of this potential is shown in
Eqs.~(\ref{eqnsphereasphereA}-\ref{eqnsphereasphereR}).  Computation
of sphere-asphere potentials is more efficient than the full RE$^2$
potential.  We take RE$^2$ parameters between the base in question and
a sphere of the LJ radius $\s$. These parameters are given in
Tables~\ref*{S1-tabsigmaP} and~\ref*{S1-tabsigmaS} for the
interactions of bases with phosphates and sugars
respectively. Base-phosphate potentials are derived from fits to the
MM3 potential in the same manner as base-base interactions, and result
in well-depths between about $0.2$ and $0.7 \kB T$, Base-sugar
interaction parameters are selected so that their corresponding
potentials are nearly purely repulsive. This results in small values
of the energy parameters, and an interaction radius of 2 \AA~(see
Tables~\ref*{S1-tabsigmaP} and~\ref*{S1-tabsigmaS}).

\subsection{Sugar-sugar and sugar-phosphate van der Waals interactions}
Sugars may interact with non-local sugar and phosphate residues that
happen to be in spatial proximity. We model their interaction with a
Lennard-Jones potential
\be
V_{LJ}(r) = 4 \epssub{LJ}  \left[ (\s/r)^{12} - (\s/r)^6 \right]
\label{eqLJ}
\ee between residues $j>i+2$. We use a well depth of $\epssub{LJ} =
0.25$ kcal/mol and a distance of $\s = 2$\AA. This potential mainly
serves to prevent steric overlap of backbone residues.

\subsection{Base-base hydrogen bonds}
\begin{figure}[!h]
\centering
      \includegraphics[width=8cm]{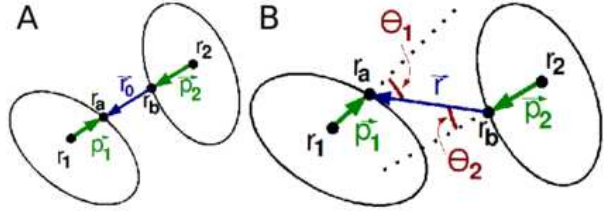}
      \caption[]{Explanation of the hydrogen bond potential
        Eq.~(\ref{eqHB}).  The left image shows two bases aligned in
        the minimum energy configuration, and the right image shows
        them in a higher energy configuration where ${\bf r} \neq {\bf
          r}_0$ and $\theta_1 \neq \theta_2 \neq 0$.}
      \label{hbondfig}
\end{figure}

Bases in DNA may pair by hydrogen bonding, an interaction not
accounted for by the RE$^2$ potential, which is based upon atomic van
der Waals interactions.  We add this in explicitly using a
phenomenological potential with a form generalized from that used in
all-atom studies~\citep{GordonDB99}: \be V^{(ij)}_{HB} = \epssub{HB}
\left[ 5 \left( {\rN \over r}\right)^{12} - 6 \left( {\rN \over r}
  \right)^{10} \right] \left( \cos^4(3 \theta_i)\cos^4(3 \phi_i) +
\cos^4(3\theta_j)\cos^4(3\phi_j) \right)
\label{eqHB}
\ee

The hydrogen bond potential for a Watson-Crick pair can be interpreted
as the sum of two potentials $V^{(i)}_{HB}+V^{(j)}_{HB}$, representing
the hydrogen bonds from base $i\rightarrow j$ and base $j\rightarrow
i$ respectively. The potential can be best understood from
Fig.~\ref{hbondfig}B. $V^{(i)}_{HB}$ is a three-body potential between
the center of mass of base $i$ (${\bf r}_i$), a point on the exterior
of the ellipsoid considered to be the origin of the hydrogen bond
(${\bf\rA}$) and a point on the other base (${\bf\rB}$) corresponding
to either donor or acceptor. The hydrogen bonding in the other
direction is similarly defined.  We take ${\bf \rA}$ and ${\bf \rB}$
to be the unique points along the line segments joining ${\bf r_1}$
and ${\bf r_2}$ which intersect the surfaces of the two ellipsoids
when the bases have the positions determined by the standard
coordinates of the B isoform structure of DNA~\citep{Arnott76}. In
Eq.~(\ref{eqHB}), $r = | \rA - \rB|$ and $\rN = | \rAN - \rBN |$ in
the B isoform structure (see Fig.~\ref{figdna}B).

The angle $\theta$ in Eq.~(\ref{eqHB}) is defined in
Fig.~\ref{hbondfig}B. The angle $\phi$ in Eq.~(\ref{eqHB}) captures
the empirical fact that multiple hydrogen bonds between bases results
in a rigidity to fluctuations that would prevent base $2$ from
rotating about ${\bf p}_2$ for example. Let ${\bf n}_1$ and {${\bf
    n}_2$} be the normals of bases $i$ and $j$ respectively. The
$\phi_i$-dependent terms breaks the symmetry of rotations of base $i$
about ${\bf p}_1=\rA -{\bf r}_1$ and rotations of $j$ about ${\bf
  p}_2$.  In Fig.~\ref{figdna}B, these normals are all pointing out of
the page for simplicity.  For the case of $\phi_i$ we project the
normal ${\bf n}_2$ onto the plane perpendicular to ${\bf p}_1$ and
$\phi$ will be the angle between this projected vector and ${\bf
  n}_1$: $\cos(\phi_i) = {\bf n}_1 \cdot \frac{{\bf n}_2 - {\bf
    n}_2\cdot {\bf p}_1}{|{\bf n}_2 - {\bf n}_2\cdot {\bf p}_1|}$. We
perform the symmetric calculation ($i\leftrightarrow j,
1\leftrightarrow 2$) to get $\phi_j$. Since the $\theta$-dependent
terms in Eq.~(\ref{eqHB}) already provide rigidity against
fluctuations such as those shown in Fig.~\ref{figdna}B (right) as well
as to buckling out of plane, we found the above formulation superior
to simply using ${\bf n}_1\cdot {\bf n}_2$ which would ``overcount''
the potential cost of those fluctuations.

As in all-atom hydrogen bond potentials, raising the cosines in
Eq.~(\ref{eqHB}) to the power of 4 ensures that the potential is
strongly orientation dependent. The factor of $3$ inside the cosine is
a result of the fact that we apply an angle cutoff of $\pi/6$ to both
$\theta$ and $\phi$. Therefore, when either angle reaches $\pi/6$, its
contribution to the hydrogen bond potential vanishes. This also
ensures a well-depth $\epssub{HB}$, and avoids multiple minima in the
cosine terms.  The equilibrium separation $\rN$ is chosen so that the
minimum of the potential matches the correct equilibrium separation in
the crystal structure.

Energy well depths must account for the fact that when a base pair
hydrogen bond is broken, the two bases can form hydrogen bonds with
the solvent. In implicit water, the well depth of the hydrogen bond
represents the free energy difference between bonded and unbonded base
pairs in solvent, which has been determined experimentally to be $1.2
\pm 0.4$ kcal/mol for each hydrogen
bond~\citep{fersht1987hbm}. Additionally, there is cooperativity in
the energies of three hydrogen bonds, which makes G-C bonds more than
1.5 times stronger than A-T bonds. The stability of a base pair is
also reduced by the presence of neighbors in the crystal structure,
which induce strain on the base pair~\citep{MallajosyulaSS05}, and
reduces the value from that of an isolated base pair~\citep{MoY06}, to
stabilities of $10.2$ kcal/mol and $17.2$ kcal/mol for A-T and G-C
pairs respectively \textit{in vacuo}. The ratio between these two
hydrogen bond well depths is $1.686$, which when applied to the
reasonable estimate of $2.4$ kcal/mol for the hydrogen bond strength
of A-T pairs in water, gives an estimate of $4.0$ kcal/mol for the
hydrogen bond strength of G-C pairs. We therefore take $2.4$ and $4.0$
kcal/mol as our well depths $\epssub{HB}$ for A-T and G-C hydrogen
pairings respectively.

We allow hydrogen bonds between any pair of bases that satisfies the
Watson-Crick pairing requirement. Due to the fact that our hydrogen
bond potential is strongly orientation and position dependent, we
found that it was very rare that a base shared a bond with more than
one base at a time: on average, about 95\% of the hydrogen bond energy
was concentrated between the putatively bonded base pairs.  The
average base-base stacking energy at typical temperatures and ionic
concentrations is approximately -1.6 kcal/mol per pair of stacked
bases. For the same conditions we find an average hydrogen bond energy
per hybridized base pair of -3.2 kcal/mol. Because a double helix of
length $N$ has $N$ base pairs and $2 N-2$ stacking interactions, the
total hydrogen bond energy and stacking are comparable in strength in
our model.

\subsection{Phosphate-phosphate electrostatic interactions}
Phosphate atoms in DNA atoms carry a partial charge close to $-e$ in
solution; in our coarse-grained model we thus assign a charge $-e$ on
each P residue. These charged residues interact by a screened Coulomb
potential in the presence of ionic solution, which is well
approximated by Debye-H\"{u}ckel theory. Ions are treated implicitly,
and the potential between charged sites is given by \be V_{qq}(r) =
{q_i q_j e^{-r/\lD} \over 4 \pi \eps_o \varepsilon r_{ij}}
\label{eqVcoulomb}
\ee
Here $\varepsilon =78$ is the dielectric constant of water, and $\lD
= (\eps_o \varepsilon \kB T/2 e^2 c)^{1/2}$ is the Debye length.
$\eps_o$ is the permittivity of free space, and $c$ is the ionic
concentration. In water with $200 mM$ ionic concentration
of KCl, the Debye length $\lD \approx 6.8$ \AA.

\subsection{Determination of local potentials by all-atom simulation}
\label{seclocalpotentials}

\begin{figure}[!h]
\centering \includegraphics[width=8cm]{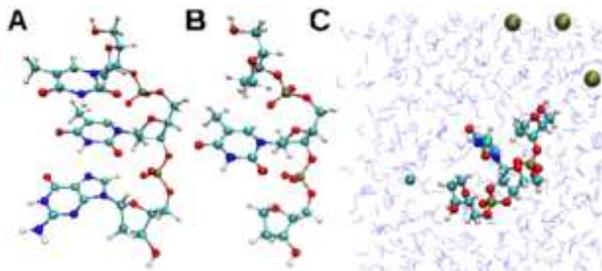}
\caption[]{Recipe for the determination of bond, angle, and dihedral
  potentials from all-atom simulations (see text).\vspace{1.0cm}}
\label{figproc}
\end{figure}

For all local potentials on the second line in Eq.~(\ref{eqV}),
no functional form is assumed {\it a priori}.  Instead the
functional form and the potential parameters are extracted from
thermal sampling of equilibrium configurations in all-atom
simulations.  For example, bond angle potentials up to an
unimportant constant are given by $\Veff(\theta) = -\kB T \ln
p(\theta)$, where $p(\theta)$ is the probability to observe a
bond angle of $\theta$. To obtain statistics that are
undistorted by the presence of potential terms already
accounted for in Eq.~(\ref{eqV}), we designed a modified system
which had no base-base interactions, and minimal Coulombic
interactions between phosphates.  Three consecutive bases are
taken with center base A, T, G, or C as in Fig.~\ref{figproc}A.
The end bases are then removed and replaced with a hydrogen of
correct bond length to each $C1'$ atom (Fig.~\ref{figproc}B).
This removes any bias in the effective potentials that would
have been due to base-base interactions, which are already
included in Eq.~(\ref{eqV}).  The molecule is then solvated in
a box of water such that the water extends $8$\AA~beyond the
boundary of the simulated molecule on all $6$ sides, and $K^+$
and $Cl^-$ ions are added at an average concentration of
$200$mM (Fig.~\ref{figproc}C). In practice this involves adding
of order $1$ ions to the system. Charge neutrality biases the
number of ions of each type: there are more positive charges
present to balance the negative phosphate backbone. The
resulting system is then simulated for $\sim 250$ns for each
base, using the NAMD simulation package~\citep{PhillipsJC05}
with CHARMM27 potential parameter set~\citep{FoloppeN00}. This
long simulation time is required for the statistics gathered to
converge within reasonable limits. Bond potentials converge to
within $1\%$ of their asymptotic value in $0.02ns$, single well
angle potentials converge in $0.5ns$, double well angle
potentials converge in $2.5ns$, and the dihedral potentials
converge in $5ns$. The longer simulation time was to be certain
of the convergence, and was not prohibitive due to the very
small size of the strand we were simulating. Dihedral
potentials converge slower: their potentials are
shallower resulting in a longer relaxation times, and a larger phase
space (larger number of atoms) is involved.

We found that the potentials extracted from all-atom
simulations often differed markedly from the commonly assumed
phenomenological forms for such potentials between the same
sets of atoms. In the supplemental materials section
\ref*{S1-secfigures}, we show representative plots of the
statistics-derived potentials for the case of bonds lengths
(Fig.~\ref*{S1-figrbond}), bond angles
(Fig.~\ref*{S1-figtheta}), and dihedral angles
(Fig.~\ref*{S1-figdihedral}).  We use A,T,G,C to represent the
location of the center of mass of each base. The parameters for
all these potentials are given in Tables
\ref*{S1-tabR}-\ref*{S1-tabimproper} in SM.

We also found that correlations between the coordinates used
for Boltzmann inversion were generally small but nonzero, with
an RMS value for cross-correlations of $\approx 0.19$. Larger
cross-correlations tended to be between overlapping degrees of
freedom such as bond $P_{3'} S$ and angle $P_{3'} S P_{5'}$. We
plot the correlation matrix in Fig.~\ref*{S1-figcorr}.

Almost all bonds show Gaussian statistics consistent with harmonic
potentials: $ V_{bond}(r) = {1\over 2} k_r (r-r_o)^2 \: ,$ however the
spring constants $k_r$ vary considerably.  The stiffest bonds were
between the center of mass of the bases and the $C1'$ sugar
residue. For this bond, pyrimidines, being the smaller bases, were
stiffer than purines.  The effective bond spring constant correlates
well with that predicted by the phonon dispersion relation for a 1-D
chain of coupled oscillators (Fig.~\ref{figkvsr}). The bond potentials
between base center of mass and $C1'$ are quite stiff, having natural
frequencies of order $0.1\ \mbox{fs}^{-1}$. Explicitly including these
would necessitate a prohibitively short time-step in coarse-grained
simulations.  For this reason we treat the base-$C1'$ system as a
single entity, i.e. the base ellipsoid plus C1$'$ residue are
hard-coded as a rigid object with fixed internal geometry. Because the
position of the neighboring sugar residue is always the same relative
to the base, a force on the sugar induces the same force as well as a
torque on the base ellipsoid.

There are $11$ different bond angles in the coarse-grained model. Five
of these have harmonic potentials, $V_{angle}(\theta) = {1\over 2}
k_{\theta} (\theta-\theta_o)^2 $, with stiffness coefficients
$k_\theta$ ranging from about $3$ to
$19\,\mbox{kcal}/\mbox{mol}\cdot\mbox{rad}^2$.  The stiffness of
S$_{5'}$P$_{5'}$S and SP$_{3'}$S$_{3'}$ are equal within error bars, and
might differ only through end effects, hence we take the average.  Six
of the bond angles were much more readily fit by double well
potentials:
\begin{eqnarray}
 V_{angle}(\theta) = -\kB T \ln \left( e^{- k_1 (\theta-\theta_1)^2/2\kB T}+A
   e^{- k_2 (\theta-\theta_2)^2/2\kB T}\right) \label{eqndoublewell}
\end{eqnarray}
with minimum angles $\theta_1$, $\theta_2$, effective stiffness
constants $k_1$, $k_2$, and dimensionless weighting factor $A$. While
typical barrier heights for interconversion were about $1 \kB T$, the
largest barrier for interconversion between wells was about $4 \kB T$,
for $\theta_{TSP_{3'}}$.  Similar to the bond potentials, the
stiffness coefficients of the harmonic bond angles show a decreasing
trend with mean distance between the participating atoms
(Fig.~\ref{figkvsrtheta}).

The minima of the parameterized angle potentials compare well
with the angles extracted from the atomic coordinates of the
standard model of (B,$10_1$,0.338)-DNA derived from the crystal
structure \citep{Arnott76} (Fig.~\ref*{S1-figangleqmVparam}) .
In cases with double-welled angle potentials, the B-form
standard model angles lie between the two parameterized wells,
closer to the deeper one. Therefore, the parameterization taken
from a piece of DNA too small to form anything like a double
helix actually biases the angle potentials to a double helix
quite well.  An exception was backbone angles, which showed a
significant discrepancy with those of the B-DNA structure.
Other potentials such as stacking may strongly influence the
equilibrium angle observed for larger strands of DNA, as noted
by Tepper and Voth~\citep{TepperHL05}, who set their
equilibrium backbone angle to $\pi$.

There are $4$ types of dihedral potential in the model: using
the notation $X$ for any of the four bases, these are
$S_{5'}P_{5'} S X$, $S_{5'}P_{5'} S P_{3'}$, $P_{5'} S P_{3'}
S_{3'}$, and $X S P_{3'} S_{3'}$ (these coordinates can be seen
in Fig.~\ref{figdna}A). The dihedral
potentials for different bases generally have different
parameters, which splits the above $4$ types into $10$ distinct
dihedral potentials. The dihedral potentials are well-fit by
the functional form 
\be 
V_{dihedral}(\phi) = \sum_{n=1}^3 K_n
(1 - \cos(n\phi- \phi_n)). 
\label{eqdihedral} 
\ee

Barriers between local minima on the dihedral potential were all less
than about $1.5 \kB T$, indicating rotations with respect to these
degrees of freedom are all quite facile. A comparison of the minima of
the resulting dihedral angles with those extracted from the crystal
structure of B-DNA can be seen in Fig.~\ref*{S1-figdihedralqmVparam}.

There is an improper angle between the normals to the planes defined
by $P_{3'}SX$ and $P_{5'}SX$, where $X$ represents the base. This
angle is taken to be zero when the $P_{3'}$ residue coincides with
$P_{5'}$, and $\pi$ when $P_{3'}SXP_{5'}$ all lie in the same
plane in the shape of a ``Y''. Generally the parameters entering into this potential were
base-dependent (see Table~\ref*{S1-tabimproper}), and well fit by
Eq.~(\ref{eqdihedral}).  There was generally a preferred angle around
$0.8 \pi$ for the purines and $\sim 0.5 \pi$ for the pyrimidines, and a
large barrier ($\sim 5\kB T$ for purines and $\sim 8 \kB T$ for the
pyrimidines) that inhibited full rotation.

We found that in the all-atom simulations bases were not free to
rotate: interactions with the rest of the molecule biased the
orientation of the effective ellipsoid to have a preferred
angle. However these interactions are already counted in base-P or P-P
interactions in Eq.~(\ref{eqV}), so to explicitly include such effects
in an angle potential representing the orientation of the base would
be redundant. Moreover we observed that explicitly adding such a
base-orientation potential derived from the all-atom statistics for 3
consecutive bases decorrelates the normal vectors of each base from
its neighbors and thus competes with stacking order.  We thus allowed
bases to freely rotate about the $C1'$-base bond.

\subsection{Langevin thermostat}
\label{sectLangevin} 
We have adapted the molecular dynamics
package LAMMPS \citep{LAMMPScite} to simulate the coarse
grained DNA model. The equations of motion are integrated using
a conventional Velocity Verlet method for translational and
rotational degrees of freedom with a timestep $\Delta t=2-10
\mbox{fs}$ depending on the degree of collapse in the system.
To avoid the well-known singularities associated with Euler
angles, a quaternion representation is used to describe the
orientation of the ellipsoids \citep{kuipers1999qar}. Since we
represent the DNA in implicit solvent, we employ a Langevin
thermostat to maintain constant temperature. The thermostat
adds forces ${\bf F}^{{\tiny (L)}}$ and torques
$\boldgreek{\tau}^{{\tiny (L)}}$ to the deterministic forces
and torques arising from the interaction potentials. In the
principal (body-centered) basis, these stochastic forces and
torques are given by 
\alpheqn
\begin{eqnarray}
F^{{\tiny (L)}}_i&=&-\zeta^{tr}_{i}v_i + \xi^{tr}_i
\label{trans-eq}\\
\tau^{{\tiny (L)}}_i&=&-\zeta^{rot}_{i}\omega_i+\xi^{rot}_i.
\label{rot-eq}
\end{eqnarray}
\reseteqn 
Here $v_i$ and $\omega_i$ denote the Cartesian components of
translational and angular velocities, $\zeta^{tr}_{i}$ and
$\zeta^{rot}_{i}$ are the eigenvalues coefficients of the
translational and rotational friction tensors and $\xi^{tr}_i$ and
$\xi^{rot}_i$ are components of white noise whose amplitudes are
related to the friction coefficients through the fluctuation
dissipation theorem.

In the body frame of the ellipsoid, the friction tensor $\zeta^{rot}$
is diagonal so that Eq.~(\ref{rot-eq}) for the torques can be
evaluated directly. The Langevin forces on the translational degrees
of freedom in the fixed lab frame, however, depend on the orientation
of the ellipsoid relative to the direction of motion. On timescales
shorter than a typical rotation time, the diffusion of an ellipsoid is
anisotropic, but crosses over to an effective isotropic diffusion at
longer times \citep{han2006bme}.  We found it easiest in calculating
these forces to first project the velocity vector into the body frame
of the ellipsoid (where the friction tensor is diagonal), evaluate
Eq.~(\ref{trans-eq}) in the body frame, and lastly rotate the
resulting force vector ${\bf F}^{lang}$ back into the frame of the
simulation cell.  The same basis transformation was performed in
calculating torques in Eq.~(\ref{rot-eq}).

The values of the translational and rotational friction coefficients
in Eqs.~(\ref{trans-eq}-\ref{rot-eq}) are taken directly from all-atom
simulations of the diffusion of a single base solvated in a (20
\AA)$^3$ box of water with 200 mMol KCl. Separate $10$ ns simulations
are performed for each base.  From these results, the effective radii
of the ellipsoid representing the base can be found by comparing the
observed friction coefficients with that predicted from continuum
hydrodynamic theory. It is encouraging that these radii are comparable
to the energetic equipotential radii at $10 \kB T$
(Fig.~\ref{figHEradii}). However it is worth noting that the radii
extracted from continuum hydrodynamics tend to be smaller than the
energetic radii, most likely due to a breakdown of the no-slip
condition at the interface between bases and water.

{\bf Energy scale for the effective simulation temperature. }  We
express temperature in units of a natural energy scale
$\eps$ in the system. We construct this energy scale by taking the
minima of the RE$^2$ potential between identical bases given as
$V_{min}$ in Table~\ref*{S1-tabsigmalike}, and averaging this value
over all bases A, C, G, T, and orientations x, y, z (an average of the
$12$ energetic values in Table~\ref*{S1-tabsigmalike}).  This results
in a value of $\epsilon = 1.45$ kcal/mol.

{\bf Sequence definitions. } Sequences used in our simulations will be
referred to with the following convention.  We will use C$_N$-G$_N$ to
mean a strand of poly(C) of length $N$ which is hydrogen bonded to its
complementary poly(G) strand of the same length. A$_N$-T$_N$ is
similarly defined. We denote by $HET$ several heterogeneous sequences,
defined in the $5'\rightarrow 3'$ direction as:
\begin{eqnarray}
	HET_{SS,25} &=& CAGGATTAATGGCGCCTACCTTACC \nonumber \\
	HET_{SS,30} &=& CATCCTCGACAATCGGAACCAGGAAGCGCC \nonumber \\
	HET_{SS,60} &=& HET_{SS,30}-CCGCAACTCTGCCGCGATCGGTGTTCGCCT \nonumber \; .
\end{eqnarray}
Finally, $HET_{DS,N}$ is a strand of $HET_{SS,N}$ that is hydrogen bonded to its complementary strand.

\section{RESULTS}

\subsection{Persistence length of ssDNA and dsDNA}

\begin{figure}[!h]
\begin{center}
\includegraphics[width=8cm]{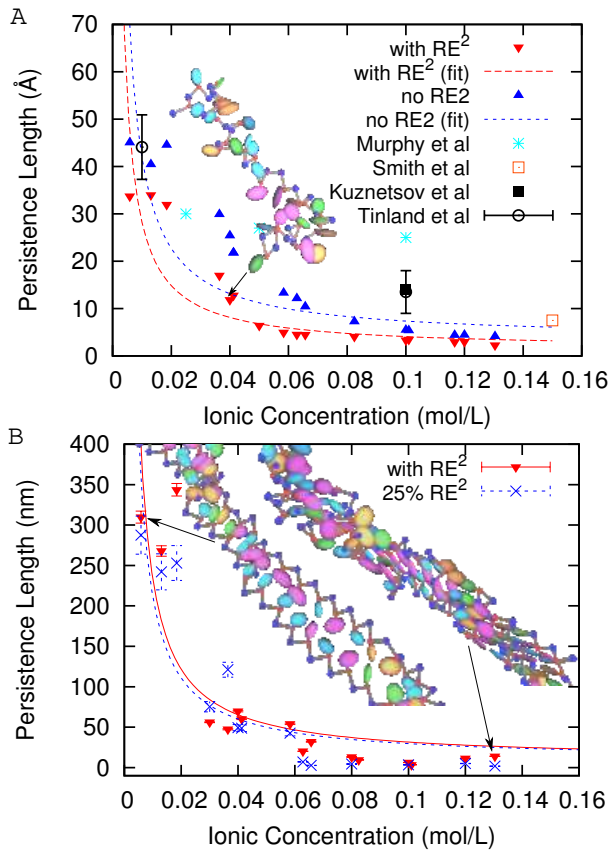}
\end{center}
\caption[]{(A) Persistence length vs ionic concentration of a random
  25bp ssDNA sequence ($HET_{SS,25}$) at a temperature of
  $0.42\epsilon$ simulated for 300 ns.  Also shown are simulation
  results for ssDNA with the RE$^2$ potential turned off (blue
  triangles).  The remaining data sets are from various experiments:
  Murphy {\it et al}~\citep{murphy2004pss}, Smith {\it et
    al}~\citep{smith1996obd}, Kuznetsov {\it et
    al}~\citep{kuznetsov2001spm} and Tinland {\it et
    al}~\citep{TinlandB._ma970381+}. (B) Persistence length vs. ionic
  concentration of a $60$bp dsDNA sequence ($HET_{DS,60}$) at a
  temperature of $0.10\epsilon$ simulated for 600 ns. Results for the
  RE$^2$ potential scaled to $1/4$ strength are also shown (blue
  Xs). Dashed lines in both panels show the theoretical model of
  Nguyen et. al.~\citep{PhysRevE.66.021801} for the persistence length
  of a polyelectrolyte, whose functional form consists of an intrinsic
  persistence length plus a term inversely proportional to ionic
  concentration. Insets show representative snapshots taken from the
  simulations (rendering with BioVEC~\citep{abrahamsson2009biovec}).
}
\label{figPL}
\end{figure}

The persistence length $\ell_p$ is a measure of the rigidity of
a polymer, and is given by the decay constant of the backbone
unit tangent vector $\hat{\textbf{t}}(s)$ as a function of base
index or position $s$ along the strand: $\langle
\hat{\textbf{t}}(s_0)\cdot \hat{\textbf{t}}(s)\rangle =
e^{-s/\ell_p}$, where $\langle\cdots\rangle$ is a thermal
average.  For ssDNA, the tangent $\textbf{t}$ was calculated by
taking the vector from the sugar (C1$'$) residue on base $i$ to
the sugar on base $i+1$, and normalizing to unity. For dsDNA,
the tangent $\textbf{t}$ was calculated by taking the vector
from the midpoint of the sugar residues of hydrogen bonded
base-pairs at $i$, to the midpoint of sugar residues of
base-pairs at $i+5$, and then normalizing to unity. The
persistence length $\ell_p$ was then obtained by exponential
fits of the data to the above correlation function. We found
that for dsDNA, if the local principal axis tangent to the
contour length of DNA (see section 3B) is used instead of the
above recipe, the same persistence length is obtained to within
$2\%$.

Without the stabilizing structure of the double helix, single
stranded DNA at ionic concentrations of 0.04 M has a
persistence length on the order 1 nm, which corresponds to 2-3
bases (the distance between successive base pairs is
approximately 0.4). $\ell_p$ is chiefly governed by the
repulsive Coulomb interaction, which tends to straighten out
the strand to maximize the distance between phosphate residues.
As the interaction becomes more screened by the presence of
ions in solution, $\ell_p$ will tend to decrease, as shown in
Fig.~\ref{figPL}. The functional dependence on concentration is
captured by a constant bare persistence length added to a term
inversely proportional to concentration
~\citep{PhysRevE.66.021801}.

The values for $\ell_p$ found in our simulations are generally less than
those observed experimentally, but it is worth noting that the
experimental measurements themselves are highly variable. This can be
understood from the persistence length being highly sensitive to other
factors such as base sequence~\citep{PhysRevLett.85.2400} and the
experimental set-up used to measure the persistence length, for
example fluorescence spectroscopy~\citep{murphy2004pss}, laser
tweezers~\citep{smith1996obd}, hairpin loops~\citep{kuznetsov2001spm}
and gel electrophoresis~\citep{TinlandB._ma970381+}.

Fig.~\ref{figPL}A also shows without the RE$^2$ potential the
stiffness of the single-stranded DNA {\it increases}. This
occurs because the RE$^2$ potential is an attractive
interaction, which tends to collapse the strand. At the
temperatures in our simulation for Fig.~\ref{figPL}A ($310K$),
the stacking and hydrogen bonds are the same order as $\kB T$.
This temperature is above the dehybridization temperature in
our model, and also above the unstacking temperature for ssDNA
described in more detail below. The consequence of this here is
that base-base interactions tend to be non-local, involving
non-consecutive bases in sequence, as can be seen from the
snapshot in Fig.~\ref{figPL}A. Thus, removing base-base
interactions (by setting the first term in eq.~(\ref{eqV}) to
zero) tends to stiffen the strand. On the other hand, at lower
temperatures the opposite behavior is observed: stiffness does
increase with increasing stacking interactions. We study this
effect in detail below.

The double helix is inherently more stable to thermal
fluctuations than a single strand, as two backbones wound
around each other provide larger elastic modulus. As can be
seen from Fig.~\ref{figPL}B, the persistence length of the
double strand has the same functional dependence on ionic
concentration as ssDNA, but is roughly $55$ times stiffer at
$0.02M$, $25$ times stiffer at $0.04M$ and $52$ times stiffer
at $0.13M$. The stiffness ratio from experimental measurements
is $\approx 66$~\citep{smith1996obd}. Weakening the RE$^2$
potential has little effect on the persistence length. That is,
due to the extra stability provided by double-stranded
hybridization, the double helix shows no collapse on the scale
of $\sim 100bp$ (see inset snapshots), so that weakening
base-base interactions only modestly reduces the stiffness due
to stacking. The effect can also be seen more evidently from
the radius of gyration (Fig.~\ref*{S1-figRGvConcentration}).

\begin{figure}[!h]
\begin{center}
\includegraphics[width=8cm]{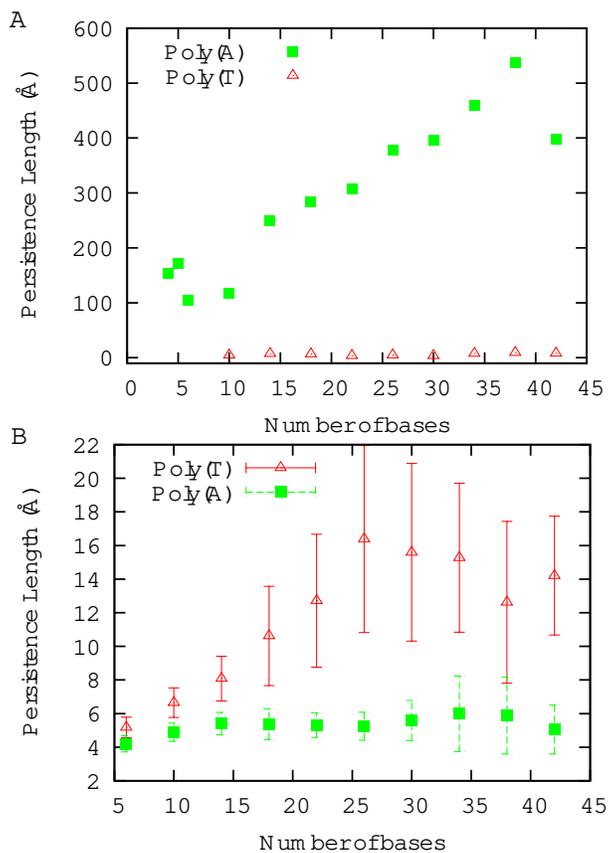}
\end{center}
\caption[]{Persistence length of ssDNA for poly(A) and poly(T) as a
  function of the number of bases in the polymer chain. Simulations
  were conducted at temperatures of (A) $0.03\epsilon$ and (B)
  $0.42\epsilon$ at ionic concentration of 40 mM for 240 ns. At these
  temperatures and ionic concentrations the distance between stacked
  bases is about $4$\AA.  Note that the temperature in (B) is above
  the hybridization temperature for dsDNA.}
\label{figsinglePLvlength}
\end{figure}

The present model predicts a larger persistence length for a
homogeneous single-strand of adenine (as large as $\sim 50$
bases or more) than the corresponding homogeneous strand of
thymine bases ($\ell_p \approx 2$ bases) at low temperatures,
as seen in Fig.~\ref{figsinglePLvlength}(A). These results are
consistent with the conclusions of Goddard \textit{et
al.}~\citep{PhysRevLett.85.2400}, who found larger enthalpic
costs for hairpin formation in poly(A) than in poly(T).
However, at high temperatures the situation is reversed, and
ss-poly(A) has a {\it smaller} $\lp$ ($\approx 1.5$ bases) than
ss-poly(T) ($\ell_p \approx 4$ bases), see
Fig.~\ref{figsinglePLvlength}(B). Adenine, being a purine, has
a stronger RE$^2$ stacking interaction (see the $z$-minima in
Table~\ref*{S1-tabsigmalike}), however all $A-A$, $A-P$
interactions are generally stronger than $T-T$, $T-P$
interactions, and at high temperature this induces a greater
degree of collapse due to non-local self interactions of DNA
strand.  The persistence length shows an increasing trend with
the length of the strand, at the temperatures and ionic
concentrations that we studied (Fig.~\ref{figsinglePLvlength} A
and B). This is due to the exaggeration of end effects on
shorter strands.  The persistence length converged to its
infinite length value for stands longer than about $7 \ell_p$
at high temperature.

\begin{figure}[!ht]
\begin{center}
\includegraphics[width=8cm]{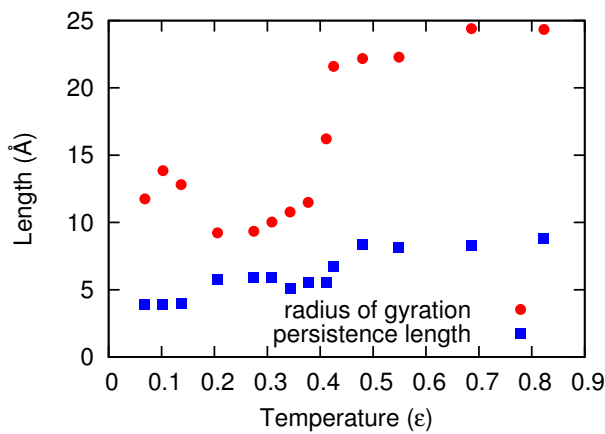}
\end{center}
\caption[]{ Persistence length and radius of gyration vs. temperature
  for ssDNA sequence $HET_{SS,25}$ at ionic concentration 0.04
  mol/L. Simulations were run for 240 ns.}
\label{figsinglePLvT}
\end{figure}

At moderate to high temperatures, we observe a heterogeneous strand to
collapse as we lower the temperature, as seen in
Fig.~\ref{figsinglePLvT}.  We used radius of gyration $R_g$ to monitor
the overall degree of collapse of the DNA, which is given by $$ R_g^2
= \langle \frac{1}{N} \sum_{k=1}^N \left( {\bf r}_k - {\bf r}_{avg}
\right)^2 \rangle,
$$ where ${\bf r}_{avg} = N^{-1} \sum_{k=1}^N {\bf r}_k$, and the
angle brackets represent the thermal average. Small values of $R_g$
correspond to collapsed states.  The general increase of persistence
length and radius of gyration in the model with temperature contrasts
with the temperature dependence of a worm-like chain ($\lp \sim
T^{-1}$). At higher temperatures thermodynamic states with larger
entropy have larger weight, the polymer expands and the
self-interactions which reduced the persistence length become less
important. The collapse temperature where the radius of gyration
suddenly increases is $\approx 0.4\epsilon$ (see
Fig.~\ref{figsinglePLvT}). A similar trend towards collapsed states
with increasing ionic concentration can be seen in
Fig.~\ref*{S1-figRGvConcentration}.

\begin{figure}[!ht]
\begin{center}
\includegraphics[width=8cm]{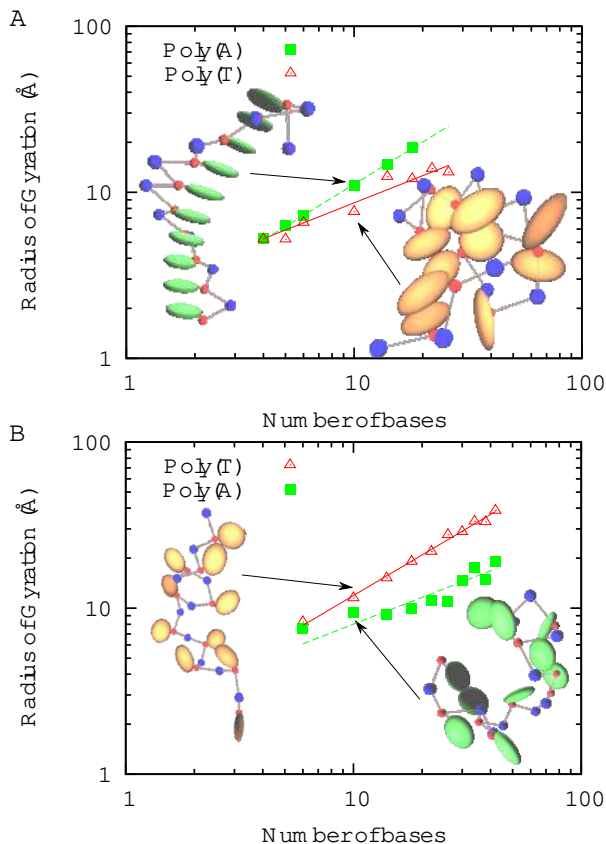}
\end{center}
\caption[]{ \label{figsinglePLvRG} Radius of gyration of ssDNA for
  poly(A) and poly(T) as a function of the number of bases obtained
  from the same set of simulations shown in
  Fig.~\ref{figsinglePLvlength} (i.e. $c=40mM$ (A) $T=0.03\epsilon$,
  (B) $T=0.42\epsilon$). $R_g$ exhibits a power law scaling
  with the number of bases with exponents of (A) $0.54\pm 0.08$ for
  poly(T), $0.84\pm 0.01$ for poly(A), and (B) $0.80\pm 0.02$ for
  poly(T) and $0.45\pm 0.07$ for poly(A), respectively.  Insets show
  representative snapshots taken from the
  simulations~\citep{abrahamsson2009biovec}.}
\end{figure}

The temperature dependence of persistence length and radius of
gyration is not a simple monotonic function. Its complexity is seen by
comparing Figs.~\ref{figsinglePLvlength} and \ref{figsinglePLvRG}. The
RE$^2$ interaction in the adenine bases is strong enough to cause a
well stacked configuration at low temperatures, whereas for thymine
bases, the stacking interaction is too weak
(Fig.~\ref{figsinglePLvRG}A). At higher temperatures,
Fig.~\ref{figsinglePLvRG}B shows that the stronger RE$^2$ interaction
causes collapse, a fact confirmed by Fig.~\ref{figPL}A.

A slightly stronger or weaker van der Waals potential between bases
results in DNA that is either collapsed or expanded respectively. Our
parameterized DNA is poised between an expanded and collapsed state
(see Fig.~\ref{figsingleRGvRE2}), so the actual state of DNA would be
highly sensitive to conditions affecting base-base interactions.  {\it
  In vivo} mechanisms for modulating base-base interactions include
DNA methylation, or potentially nucleosome post-translational
modifications utilized for gene regulation such as histone
phosphorylation, acetylation, or methylation~\citep{Henikoff08}.

\subsection{Twist and Stacking of dsDNA}
\begin{figure}[!h]
\begin{center}
\includegraphics[width=8cm]{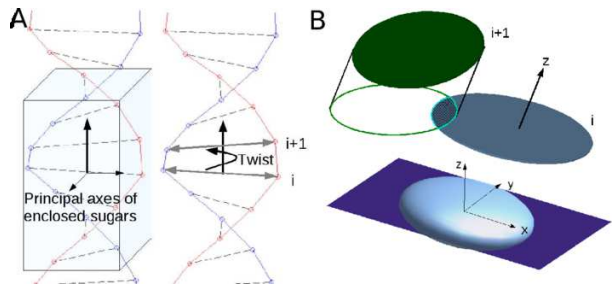}
\end{center}
\caption[]{(A) Calculation of local helical axis and twist (see
  text). (B) Calculation of the degree of stacking, or stacking
  fraction.  Principal axes are found for each ellipse and a plane
  normal to the $z$ axis is taken through the center of the base as
  shown in the bottom figure. This defines an ellipse for each
  base. The stacking is defined by projecting ellipse $i+1$ onto
  ellipse $i$ and {\it vice versa}, and measuring the overlap (see
  text). The top of (B) shows a visualization of this projection.  }
\label{fighelixstack}
\end{figure}

The twist is defined as the average angle that the backbone of the
double helix rotates about the helical axis for each successive base
pair. Fig.~\ref{fighelixstack}A visualizes the calculation of the
local twist at position $i$ along the DNA.  To obtain the helical axis
at a given position along the double helix, we take the positions of
the two sugars opposite the hydrogen-bonded bases at that position, as
well as the sugar pairs up to three bases above and below that
position. From the sugar coordinates, we compute the principal moments
of inertia, and take the moment of least rotational inertia to be the
helical axis. This method fails if the double helix persistence length
drops to be on the order of three base pairs or if the helix
dehybridizes. However, neither of these scenarios occur for simulation
parameters giving a stable double-helix.  The sugar-sugar vector
rotates around the helical axis as one proceeds along the bases; the
angle between the $i$th and $i+1$th sugar-sugar vectors is the {\it
  local} twist. This quantity is then averaged along the strand as
well as over time.  Snapshots during the simulation give the average
twist reported in the figures below.  We can similarly define the
pitch of the DNA as $\mbox{pitch} = 2\pi/\mbox{twist}$, i.e.  the
number of base pairs one must traverse for a full revolution of the
helix. Note that the observed crystal structure value for the pitch of
B-DNA is 10 \citep{Arnott76}, giving an expected twist of
$36\degrees$.

We also develop an order parameter to define how well a given
base is stacked to its neighbors. Taking the dot product of $z$
principal axes of the ellipsoids has translational symmetry in
the $x-y$ plane of either ellipsoid, and so does not capture
the concept of stacking. The method we employ instead uses area
projections, depicted in Fig.~\ref{fighelixstack}B. For each
ellipsoid, we take its cross-section of the ellipsoid in the
$x-y$ plane of its own principal axes, which is an ellipse. To
calculate the stacking fraction between bases $i$ and $i+1$, we
project cross-section $i+1$ down onto cross-section $i$. We
then take the average of this value with the equivalent
projection of $i$ onto $i+1$ in order that our definition of
stacking be symmetric. We divide the projected area by the area
that would be obtained from the B-DNA crystal structure (a
stacking fraction of $0.6$) to properly normalize.  Because of
twist, bases are not perfectly stacked in the crystal
structure, so it is possible to see stacking fractions greater
than $1$. This definition of stacking mathematically represents
the intuitive notion of stacking very well, namely the degree
to which the flat parts of the two objects overlap their areas.
The stacking fraction is a geometrical order parameter, but
correlates strongly with the RE$^2$ energies between
neighboring bases on either of the two strands of the DNA (see
Fig.~\ref*{S1-figstackvsenergy}). Thus stacking fraction
accurately captures the base-base van der Waals stacking
energies in addition to quantifying the structural features of
DNA.

\begin{figure}[!h]
\begin{center}
\includegraphics[width=8cm]{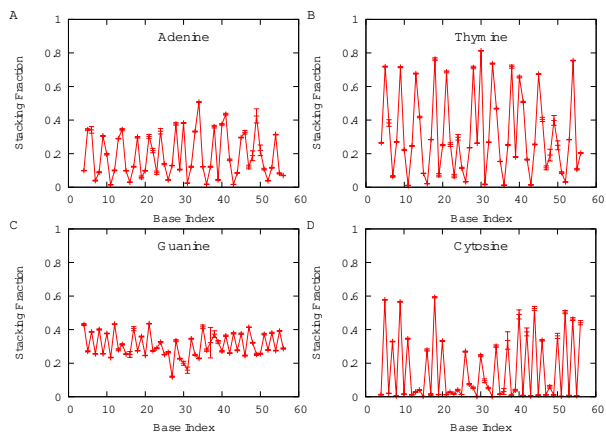}
\end{center}
\caption[]{Time averaged stacking fractions of homogeneous strands in
  a helix with their complementary strand. Figs. A and B are obtained
  from simulations of 60 base dsDNA, consisting of a homogeneous
  strand of poly(A) hydrogen bonded to a similar strand of
  poly(T). The figures show the stacking of the adenine bases with
  themselves and the thymine bases with themselves
  respectively. Figs. C and D are obtained similarly to A and B except
  with guanine and cytosine bases. All simulations were conducted at
  $T=0.07\epsilon$ and an ionic concentration of 0.04 mol/L. The total
  simulation time is 800ns. The average stacking fraction for each of
  these simulations is $0.1889 \pm 0.0185$ for adenine, $0.3178 \pm
  0.0343$ for thymine, $0.3065 \pm 0.0099$ for guanine and $0.1593 \pm
  0.0271$ for cytosine.}
\label{stacktimeav}
\end{figure}

\begin{figure}[!h]
\begin{center}
\includegraphics[width=8cm]{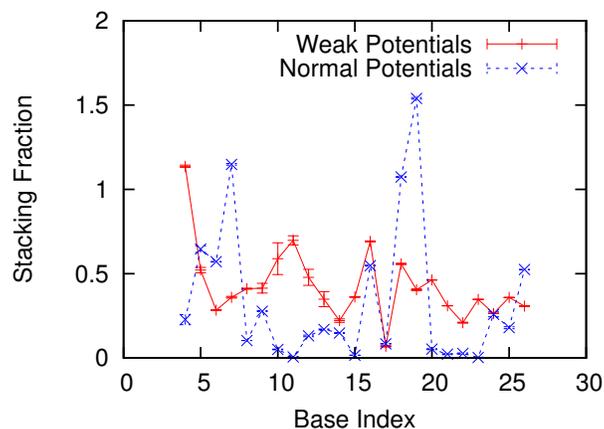}
\end{center}
\caption[]{Time averaged stacking fraction of the bases along a strand
  for both the all-atom parameterized potentials and another
  simulation with weakened parameters. Here the bond potentials were
  weakened by a factor of 0.5, the angle potentials by 0.1 and the
  dihedral potentials were disabled. This indicates that the decimated
  stacking pattern we observe for bases is due to frustration between
  stacking interactions and the other potentials in the model (bond,
  angle, dihedral).  These simulations were done with $HET_{DS,30}$ at
  an ionic concentration of 0.04 mol/L and a temperature of $0.07
  \epsilon$.}
\label{figstacklownormspring}
\end{figure}

We found base stacking to be very heterogeneous, with traces of
periodicity along the strand having a period of roughly four to five
bases (see Fig.~\ref{stacktimeav}). Bases seem to stack well in small
groups, at the expense of poorer stacking in bases very nearby along
the strand. This results in kinks in the stacking structure of the
strand, with the distance between kinks being only a few base
pairs. An interesting trend is that the purines, the larger bases with
greater stacking interactions, fluctuate far less and show more
consistent stacking.  On the other hand pyrimidines show larger
extremes in their decimated stacking pattern, stacking intermittently
more strongly than purines, and completely unstacking.  The stronger
stacking interactions of the purines apparently induce convergence to
the average value. The decimation pattern that we observe for the
bases is due to frustration between stacking interactions and the
other potentials (bond, angle, dihedral), and reducing these other
potentials resulted in less fluctuation, and a larger average stacking
(see Fig.~\ref{figstacklownormspring}). This situation is reminiscent
of a Frenkel-Kontorova model where a competition between two
incommensurate length scales corresponding to the equilibrium
separation of a 1D chain of oscillators and a periodic underlying
potential results in frustration-induced domains of oscillators
\citep{aubry8sac}.

To understand which parts of the Hamiltonian control the large scale
properties of twist and stacking of dsDNA, we applied global
multiplicative scaling factors to the energies of individual potential
classes, such as RE$^2$, bond, angles, dihedrals and hydrogen bonds.
Our simulations show that the potential most sensitively affecting the
structure of the double helix is the base-base RE$^2$
interaction. Increasing the base-base RE$^2$ potential reduces the
contour length of the helix, (see Fig.~\ref*{S1-fig_contour}), increases
the stacking fraction, and tightens the twist over a large range of
interaction strength.

\begin{figure}[!h]
\begin{center}
\includegraphics[width=8cm]{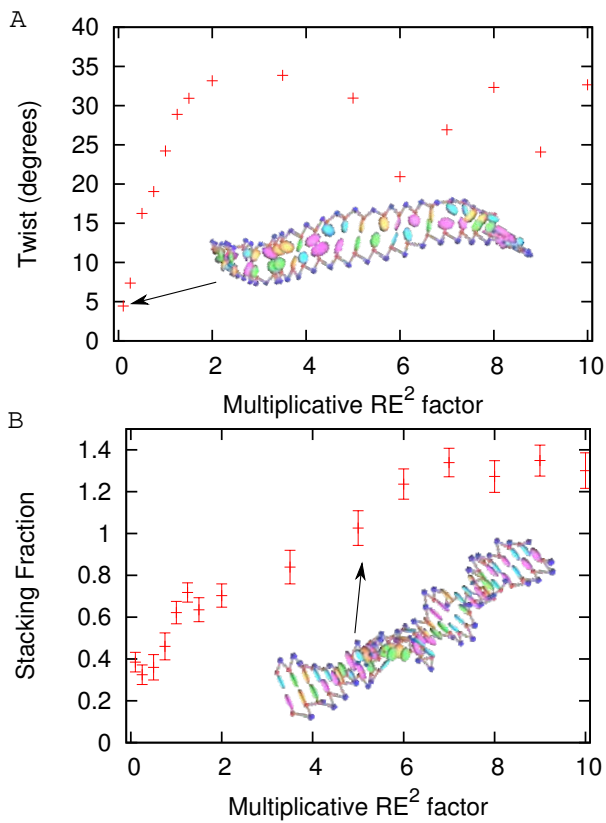}
\end{center}
\caption[]{Effect of scaling the RE$^2$ energies on (A) twist and (B)
  stacking of double stranded DNA for a random sequence
  $(HET_{DS,30})$ at zero temperature and $40$ mM ionic
  concentration. Simulations were started from the standard model
  (B,$10_1$,0.338)-DNA structure\citep{Arnott76} and allowed to
  equilibrate by energy minimization
  to structures having the values shown at $T=0$. The equilibration
  time was about $80$ns, implying shallow energetic gradients of
  collective modes. Insets show representative snapshots taken from
  the simulations~\citep{abrahamsson2009biovec}.  }
\label{figstackpitchre2}
\end{figure}

\begin{figure}[!h]
\centering
      \includegraphics*[width=8cm]{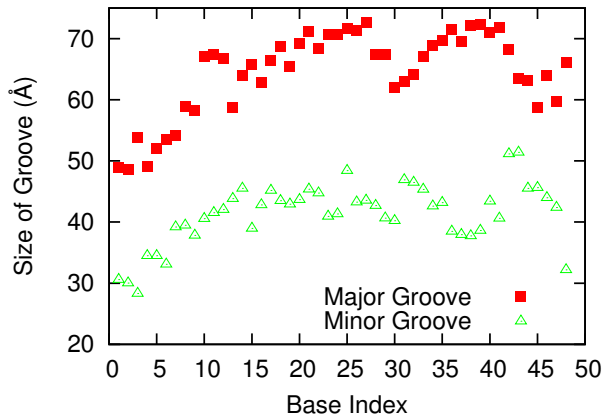}
\caption[]{The size of the major and minor grooves as a function of
  position along the double helix. This is a finite temperature
  simulation of ionic concentration 0.04 mol/L as in
  Fig.~\ref{fighelixstackconc}. The errors are of the order of the
  symbol size. The double helix makes about two turns over the length
  of the strand. }
      \label{figgroove}
\end{figure}

Perhaps surprisingly, we found that the base-base attraction also
appears to induce twist of the double helix. A picture that has
emerged from computational models of DNA structure
\citep{PhysRevE.68.021911} is that twist results from the competition
between electrostatic repulsion of phosphate groups and favorable base
stacking interactions, with stronger stacking interactions favoring
alignment of the bases and thus putatively tending to straighten the
helix. Indeed, the van der Waals-like RE$^2$ potential is minimized
when the relative twist between stacked bases is zero.
Fig.~\ref{figstackpitchre2}A plots the helical twist as a function of
RE$^2$ interaction strength which modulates stacking
interactions. This clearly shows an increase in twist with
stacking interaction strength over a large range of the RE$^2$
interaction. These results were obtained by a zero 
temperature energy minimization, starting from the expected crystal
structure of B-DNA. Finite temperature reduces the overall values but
does not change the trend. During the simulations, the helical twist
relaxed to a degree determined by the overall strength of the RE$^2$
potentials. At the all-atom parameterized values the twist was
$21.2\degrees$, whereas in the crystal structure it is
$36.0\degrees$. Despite this quantitative discrepancy we did find that
the potential function~(Eq.~(\ref{eqV})) reproduced a double-stranded
structure with major and minor grooves (Fig.~\ref{figgroove}). The
ratio of the sizes of the grooves minor to major in the coarse-grained
model was $0.64$, as compared to the experimental number of
$0.54$. Major and minor grooves persist so long as the twist of the
DNA is $\gtrsim 10\degrees$.

\begin{figure}[!h]
\begin{center}
\includegraphics[width=8cm]{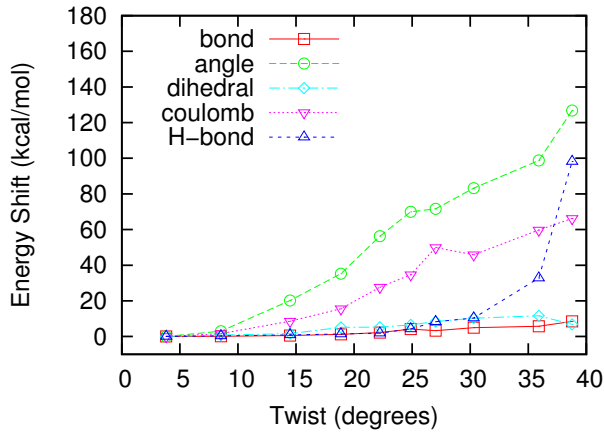}
\end{center}
\caption[]{Behavior of energy scales as functions of helical
twist, as induced by increasing stacking interactions.  Data
plotted here are taken from the same simulations as in
Fig.~\ref{figstackpitchre2}. Energies have been shifted to start from zero
to depict the relative effects of increasing twist more clearly. }
\label{figpitchenergy}
\end{figure}

The ladder conformation is preferred when the RE$^2$ potential
is scaled down to zero, where Coulombic interactions are
competing solely with local bond, angle, and dihedral
potentials. The twisted conformations of the helix bring the
phosphate residues closer together rather than farther apart
(see Fig.~\ref{figmeanPhosphateTwist}).  The direct compression
of a ladder configuration would also increase the Coulomb
energy in inverse proportion to the contour length, as well as
frustrating local potentials such as angular potentials (see
Fig.~\ref{figpitchenergy} and description below). To avoid this
energetic cost as DNA is compressed, the system can lower its
energy by structurally relaxing into a helical conformation.
Upon helix formation, the potential energy of terms such as
angles and coulomb lower more than the RE$^2$ potential energy
raises due to shearing the stacking pattern. Put another way,
the stacking interactions do not favor the twisted
conformation, they favor proximity of the bases: the total
RE$^2$ energy in a helical B-DNA conformation of $HET_{DS,30}$
is about $-205$ kcal/mol, while if the twist is set to zero
from this conformation by forcing a ladder initial condition,
the RE$^2$ energy is $-270$ kcal/mol before relaxation. However, the
other potentials favor helix formation more so than
maintaining the ladder, and so break translational symmetry
along the DNA contour when forced into proximity by the RE$^2$
potential. The correlation between RE$^2$ energy and stacking
geometry (see Fig.~\ref*{S1-figstackvsenergy}) along with the competition
between potentials in the system together imply that as the strength
of the RE$^2$ 
interaction continues to increase, the system must eventually favor
ladder-like configurations as bases are more properly stacked. 
This can be seen in figures~\ref{figstackpitchre2}A,B in the range of
multiplicative RE$^2$ factor from $5$ to $10$. The twist begins to
decrease (albeit with large scatter in the equilibration data) as the
stacking fraction continues to monotonically increase above values
present in the crystal structure.

The dominant interaction governing the stacking fraction is the
also strength of the base-base RE$^2$ potential. There was
significant relaxation of the B-DNA model structure at the
all-atom parameterization values. The twist and stacking
fraction obtain their B-DNA model values only when base-base
interactions are magnified by a factor of $5$, as can be seen
from Fig.~\ref{figstackpitchre2}B. This may indicate that
cooperative many-body effects beyond the superposition of
pairwise Lennard-Jones interactions are governing the stacking
interactions between bases.  It is worth noting that large
dynamical fluctuations in the DNA structure are also observed
in all-atom simulations~\citep{levitt1983csd}.

Increasing the stacking energy (via the RE$^2$ potential) increases
the twist, and this effect frustrates nearly all other interaction
energies in the system.  Fig.~\ref{figpitchenergy} plots the other $5$
contributions to the energy as a function of the induced helical twist
due to increasing the RE$^2$ energy. The values of the twist were
taken from the monotonic relationship in
Fig.~\ref{figstackpitchre2}A. We found RE$^2$ energy to be the optimal
parameter to vary the helicity in the subsequent simulations.  The
increased twist of the double helix frustrates the Coulomb and angle
potentials most significantly. As the twist of the helix increases,
the energy of the angle potentials increases, indicating that the
all-atom parameterized angle potentials favor a ladder configuration.
Thus the base-base stacking interactions, and not the angle
potentials, govern the tendency to helix formation, a result in
agreement with previous coarse-grained studies~\citep{TepperHL05}.

\subsection{Increasing temperature reduces twist and stacking in dsDNA in a sequence-dependent fashion}

\begin{figure}[!h]
\begin{center}
\includegraphics[width=8cm]{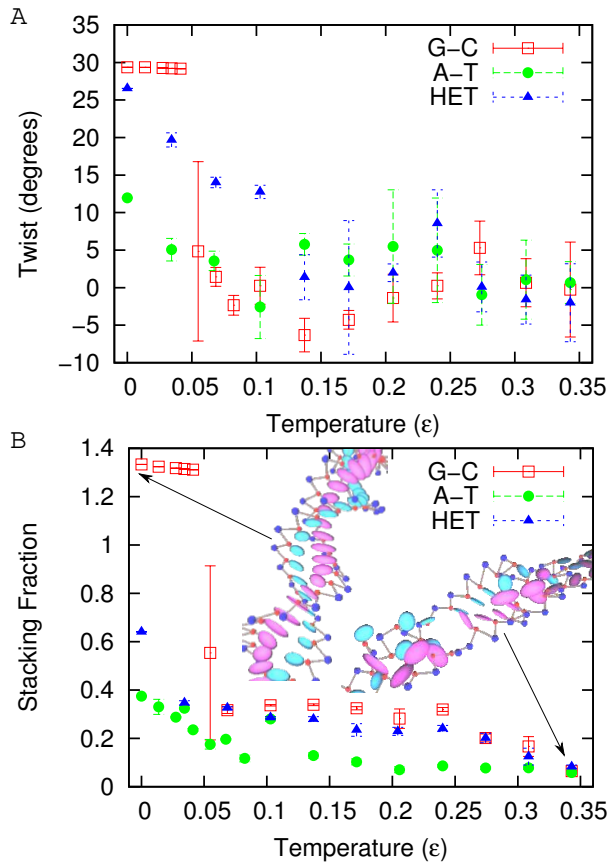}
\end{center}
\caption[]{Behavior of twist and stacking versus temperature. This is
  a simulation of sequences G$_{60}$-C$_{60}$, A$_{60}$-T$_{60}$, and
  $HET_{DS,60}$ simulated at an ionic concentration of 0.04 mol/L for
  800 ns.  Insets show representative snapshots taken from the
  simulations \citep{abrahamsson2009biovec}.  }
\label{fighelixstacktemp}
\end{figure}

Fig.~\ref{fighelixstacktemp}A shows the twist as a function of
temperature. The sequence $G_{60}$-$C_{60}$ maintains a twist of
approximately $29\degrees$ until $T = 0.05 \epsilon$ at which point it
drops sharply in a manner suggestive of a phase transition,
accompanied by large fluctuations manifested by a sudden increase in
the statistical error at $T = 0.05 \epsilon$. The helicity is
approximately zero for higher temperatures.  The stacking fraction for
$G_{60}$-$C_{60}$ shows similar behavior than the twist
(Fig.~\ref{fighelixstacktemp}B), with phase transition around $T =
0.05 \epsilon$. This transition involves a reduction of order within
the double-stranded structure, and is distinct from dehybridization
(the $HET_{DS,60}$ double strand was observed to dehybridize at a
temperature around $0.31 \epsilon$, and took about $330$ ns to
dehybridize completely). Above the ``untwisting'' transition
temperature, the stacking is still appreciable, although the twist
does not appear to be significant.  We found that $A_{60}$-$T_{60}$
maintains very little twist in our simulations. At zero temperature it
has a weak twist of $12\degrees$. This sequence likewise showed the
least stacking order among the sequences we studied. The sequence
$HET_{DS,60}$ has much smoother behavior with changing temperature. At
zero temperature, it has a twist of $26\degrees$, slightly less than
homogeneous G-C DNA, and intermediate stacking fraction to G-C and
A-T. The phase transition behavior present in $G_{60}$-$C_{60}$ is
smoothed out for the heterogeneous sequence, yet it curiously
maintains helicity until a higher temperature. Stacking is
intermediate to G-C and A-T sequences at all temperatures, and also
has a broadened transition.

\subsection{Coulomb interactions oppose both twist and stacking in dsDNA}

\begin{figure}[!h]
\begin{center}
\includegraphics[width=8cm]{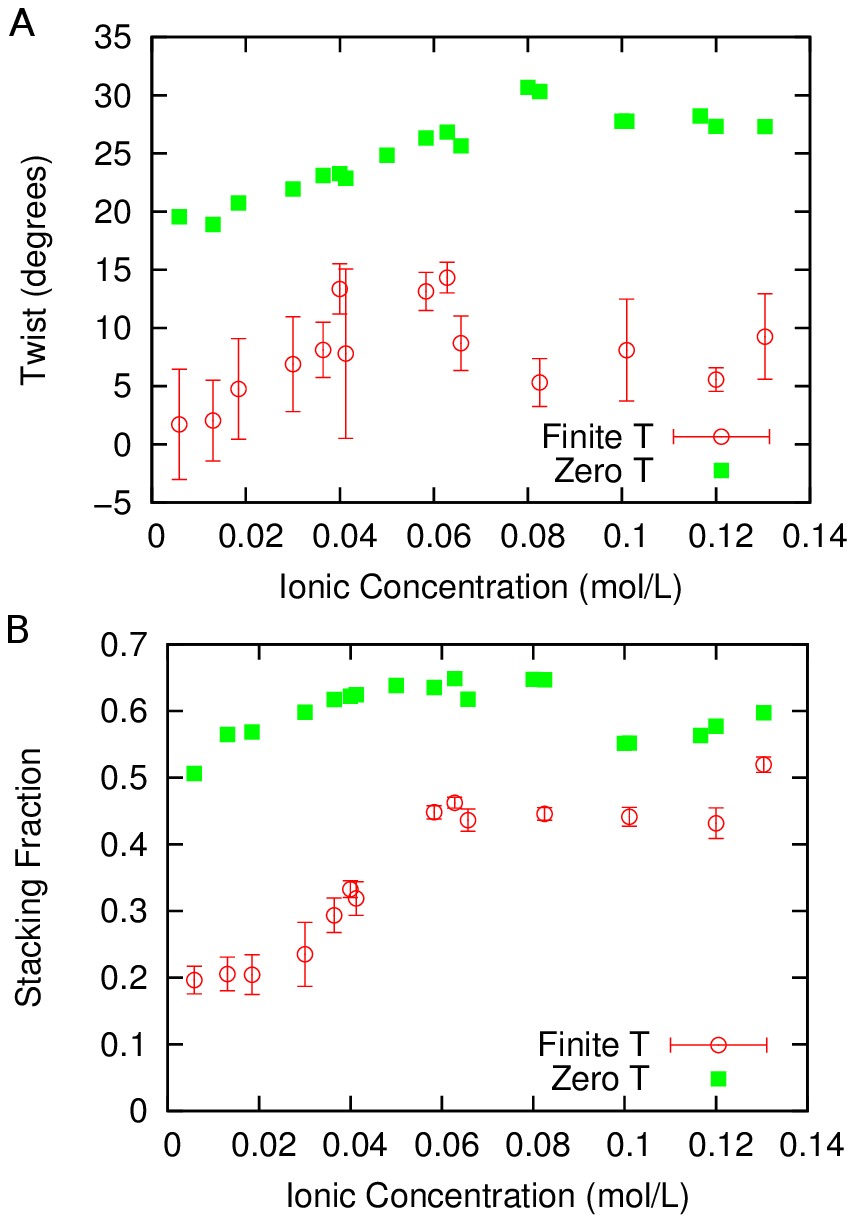}
\end{center}
\caption[]{Twist and stacking of double stranded DNA {\it vs} ionic
  concentration.  Thermal averages at finite $T$ are taken from a
  600ns simulation of $HET_{DS,60}$ at $T=0.10\eps$.  Zero temperature
  data is taken from an $80$ns simulation of $HET_{DS,60}$, and data
  from the equilibrated state is shown.}
\label{fighelixstackconc}
\end{figure}

Fig.~\ref{fighelixstackconc} shows the twist and stacking of the
$HET_{DS,60}$ sequence as a function of ionic
concentration. Increasing ionic concentration reduces the repulsive
Coulomb energy by decreasing the Debye length in
Eq.~(\ref{eqVcoulomb}).  As phosphate backbone charges are more
effectively screened in stronger ionic solution, the RE$^2$ potential
starts to dominate over the Coulomb potential, which reduces the
contour length. To minimize frustration of the other potentials upon
this compression, the helix twists to compensate. Thus an increase in
ionic concentration increases the helical twist in an indirect way
(see Fig.~\ref{fighelixstackconc}A).  The effect saturates as the
ionic concentration is increased, whereupon other potentials such as
angle and dihedral eventually dominate over the Coulomb potential.

Fig.~\ref{fighelixstackconc}B shows that the stacking fraction
increases with ionic concentration. This can be understood from the
interpretation of Fig.~\ref{figpitchenergy}, where it is seen that the
Coulomb energy opposes stacking, so that ameliorating it by increasing
ionic concentration would increase stacking propensity. As ionic
strength is increased, the DNA strand compresses due to the weakening
of Coulomb repulsion. Thermal motion is less
effective at eliminating stacking in this more compressed state, due
to the steric constraints manifested in the RE$^2$ potential.
Finite and zero temperature simulations in
Fig.~\ref{fighelixstackconc}  obey the same trend.

\subsection{The model shows chiral preference for the right-handed helix}
To address the question of whether the present physicochemical
based model exhibits chirality, we performed finite temperature
simulations of $HET_{DS,30}$ DNA starting from two different
initial conditions: one from the (right-handed) B-DNA standard
structure, and another from a structure with the azimuthal
angle between successive base pairs reversed from $+36\degrees$
to $-36\degrees$. The latter gives an "anti-B-form" DNA with
left-handed helix as the initial condition. Each configuration
was observed to relax to a metastable equilibrium conformation
for the duration of the simulation, analogous to either B-form
or Z-form DNA.

\begin{figure}[!h]
\begin{center}
\includegraphics[width=8cm]{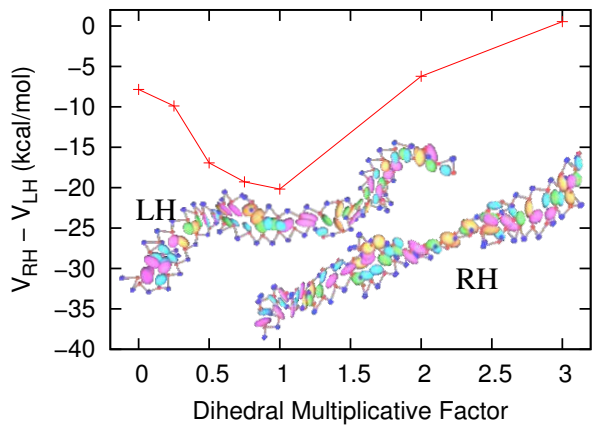}\\
\end{center}
\caption[]{Difference in the thermal average potential energy between right-handed
and left-handed forms of ds-DNA (sequence $HET_{DS,30}$), as a function of dihedral strength.
Thermal averages are taken from
  160ns simulations at $T=0.03\eps$. Inset images show
  representative snapshots from the simulations at dihedral multiplicative factor
  unity~\citep{abrahamsson2009biovec}. }
\label{figLHRHvsDihedral}
\end{figure}

The difference of the thermal average potential energy
(Eq.~(\ref{eqV})) between the right-handed and left-handed
forms of DNA is plotted in Fig.~\ref{figLHRHvsDihedral} as a
function of the strength of the dihedral potential. The
strength of the dihedral potentials are all simultaneously
varied by adjusting an overall multiplicative factor. We chose
to vary the dihedral potential because it is an obvious chiral
term in the model's potential function. Specifically, the
following dihedral potentials show bias toward the right-handed
(B-DNA) helix (with the strength of the bias given in brackets
in units of kcal/mol): S$_{5'}$P$_{5'}$SP$_{3'}$ (0.25),
P$_{5'}$SP$_{3'}$S$_{3'}$ (0.06), S$_{5'}$P$_{5'}$SA (0.40),
S$_{5'}$P$_{5'}$SC (0.18), S$_{5'}$P$_{5'}$SG (0.62). The
potential S$_{5'}$P$_{5'}$ST biases toward a left-handed
(Z-DNA) helix, but at a strength of only 0.03 kcal/mol. The
remaining potentials are chiral symmetric.

The model shows an energetic bias towards the right-handed
B-form of DNA, which at the temperature of the simulation was
about 0.67 kcal/mol$\cdot$base pair, for values of the dihedral
potential obtained from all-atom parameterization.
Interestingly, the chiral bias towards right-handed DNA is
maximal at this value of dihedral strength.
Fig.~\ref{figLHRHvsDihedral} also shows a chiral preference
even in the absence of dihedral potentials. This is evidence
that the origin of handedness in the model results from a
coupling between potentials that individually are achiral, but
collectively these potentials result in chiral constituents
that when coupled together (e.g. by stacking interactions)
yield a preference for the right-handed helix. We elaborate on this
further in the discussion section, but note for now that the structure
of the Phosphate-Sugar-Base moieties which constitute the building
blocks of DNA are themselves achiral objects by virtue of their
absence of a center of inversion or mirror plane. 

Experimental measurements of the preference of B-DNA over Z-DNA
give a free energetic difference of about $0.33$
kcal/mol$\cdot$bp~\citep{PeckLJ83}. The above energetic
difference of $0.67$ kcal/mol$\cdot$bp in the computational
model does not account for entropic differences between the B
and Z forms. If one takes the magnitude of the value in the
model seriously it would imply that the Z form has larger
entropy than the B form.

As is evident from the snapshots in the inset of
Fig.~\ref{figLHRHvsDihedral}, the stiffness of Z-DNA was seen
to be softer than that of the B-form, with persistence length
$\lp$ reduced by about a factor of $2$ at the all-atom
parameterized values. Electron microscopy measurements of chain
flexibility in Z-DNA have shown moderate increases of
persistence length $\lp$ of about $30\%$~\citep{RevetB84},
however these measurements were for poly(dG-dC) that had been
adsorbed onto a 2-D surface which could introduce new
interactions modifying $\lp$ . Since a larger $\lp$ must
correspond to stronger stabilizing interactions whose effect
would be to increase the bending modulus, one would expect from
energetic arguments that the thermodynamically less stable
Z-form of DNA would have a shorter persistence length, as we
observed in our simulations. On the other hand the increased
number of base pairs per turn in the crystal structures implies
a larger stacking fraction in Z-DNA, however only by about
10\%. Moreover the rise per base pair is almost 50\% larger in
Z-DNA, implying reduced base-stacking interactions which would
also reduce the stiffness. Reconciling the computational and
experimental observations of persistence length of Z-DNA is a
topic for future study.

\section{CONCLUSIONS AND DISCUSSION}
In this paper, we have introduced a coarse-grained model of
DNA, using rigid-body ellipsoids to model the stereochemistry
of bases. This model captures the steric effects of
base-stacking, the stability of base-pairing hydrogen bonds,
screened Coulombic repulsion of the phosphate backbone,
nonlocal interactions between base-base and base-backbone, as
well as backbone rigidity due to bond, angle and dihedral
potentials. Local effective potentials along the backbone are
obtained from the statistics of all-atom simulations in
explicit solvent. Base-base and base-backbone interactions are
obtained from best fit between van der Waals interactions in an
all-atom model and an anisotropic potential between effective
ellipsoids~\citep{BabadiM06}. Hydrogen bonds are modeled by
adapting a functional form used in all-atom simulations to
ellipsoidal bases, and phosphate-phosphate interactions are
modeled through a mean-field screened Coulomb potential with
Debye length dependent on ionic concentration.

The radii along the principal axes of the bases as defined by
equipotentials of the anisotropic energy function correlate
well with the base radii as determined from hydrodynamic
diffusion measurements in all-atom simulations. The stiffness
constants in the bond and angle potentials correlate well with
that predicted by the phonon dispersion relation for a 1-D
chain of harmonic oscillators.

The model is physico-chemistry based and uses no structural
information (i.e. no G\={o} potentials~\citep{Ueda75}) to provide bias
towards the DNA crystal structure. The potentials result in a stable
double-stranded helix with both major and minor grooves, and a
persistence length for single- and double-stranded DNA comparable to
experimental values.

We have introduced quantitative recipes appropriate for the dynamical
trajectories in molecular dynamics to quantify structural order
parameters in the system such as degree of stacking and amount of
twist. We investigated these structural properties along with other
quantities such as persistence length, radius of gyration, and
chirality, for both single- and double-stranded DNA where appropriate,
as various environmental factors such as temperature and ionic
concentration were varied. We also investigated structural order in
ss- and ds-DNA as internal parameters such as number of bases, base
sequence, and stacking strength were varied.

We find that at lower temperatures in the model, ss poly(A)
stacks significantly more strongly than ss poly(T), consistent
with the conclusions from the Libchaber
group~\citep{PhysRevLett.85.2400}. However, at higher
temperatures another regime is reached where non-local
interactions between bases govern the persistence length:
poly(A) forms a collapsed globule with shorter persistence
length than poly(T) which forms a more expanded globule. The
scaling exponents for the radius of gyration likewise show
inverse behaviors in these two temperature regimes. The
persistence length of ss-DNA initially decreases with
increasing temperature in accord with the worm-like chain
model, however at higher temperatures where non-local
interactions become important, the persistence length shows an
increasing trend over a large range of temperature, while the
radius of gyration of the DNA globule expands through a
collapse-transition temperature. In other words, below the
crossover temperature, stacking interactions stiffen the chain
and increase the persistence length, while above it nonlocal
base-base van der Waals interactions (which are inseparable
from stacking interactions) soften the chain and decrease
$\lp$.

We also investigated the interplay of forces that results in twist for
ds-DNA. We find that under typical conditions, base-stacking
interactions are the dominant 
factor in driving twist, in spite of the fact that a ladder
configuration would minimize base-stacking energy.  Increasing the
putative base-stacking strength frustrates all other interactions in
the system as twist increases, indicating no other interaction could
be responsible for inducing twist.  However, the RE$^2$ potential responsible
for base-stacking is achiral, and  
energetically minimized when bases are stacked directly on top of each
other. We thus infer that base stacking enhances the chiral 
properties of the constituent components in DNA by
bringing them in close proximity, resulting in increased twist.  
Moreover, when base interactions are sufficiently strong ($\sim
5$-$10X$ their putative value), bases eventually stack more directly
on top of each other at the expense of twist. This results in
non-monotonic behavior of the twist as a function of stacking
strength. Both
twist and stacking increase as Coulomb interactions are more
effectively screened.  Even in the native
thermodynamically stable structure, DNA is under stress and thus
strained due to competition between the various potentials.

In our model, the structure in hybridized poly(G)-poly(C) shows
different temperature dependence than poly(A)-poly(T), with
G$_{60}$-C$_{60}$ showing much more order as temperature is raised,
along with a sudden first-order-like drop in twist and stacking at a
transition temperature below the dehybridization temperature. The
untwisting with temperature is smoothed out for the heterogeneous
sequence, in a similar manner to disorder-induced broadening of a
phase transition. It is notable how sensitive the qualitative behavior
is to the sequence: A-T bases differ only mildly from their G-C
counterparts in size and energy scales, yet these differences are
enough to determine the presence or absence of critical behavior with
respect to the order parameters of helical twist and base pair
stacking.

Base stacking was analyzed at the level of resolution of individual
bases, where stacking was found to exhibit an intermittent, decimated
pattern wherein roughly four to five bases stack well in groups at the
expense of poor stacking in nearby bases. This quasi-periodic
structure is reminiscent of systems frustrated by incommensurate
length scales such as the Frenkel-Kontorova model. Consistent with
this notion, decreasing the putative strength of bond, angle, and
dihedral interactions in the model resulted in less stacking
heterogeneity, and an increase in the degree of stacking. The
heterogeneous stacking pattern was observed over the total simulation
time of about a $\mu s$. One would anticipate that over longer time
scales the specific heterogeneous pattern would shift to other
metastable configurations.

Including anisotropic van der Waals interactions through the RE$^2$
potential introduces a large number of parameters in the model, so
that the reduction in total number of parameters from that required in
all-atom simulations is not dramatic. For each of the 10 base-base
interactions, there are $6$ radii, $6$ well-depths, an overall energy
scale $A_{12}$ and an atomic length scale $\sigma_c$, for a total of
$140$ parameters.  Base-backbone potentials introduce another $80$
parameters.  Including masses, bonds, angles, dihedrals, screened
electrostatics, Langevin coefficients, phosphate charge, and
temperature gives a total of $382$ parameters. To simulate the same
system using all atom potentials requires at least $600$ parameters
including van der Waals parameters, bonds, angles, dihedrals, and
atomic properties such as mass, charge and diffusion coefficient (this
number can increase to thousands if accurate van der Waals potentials
are sought).

On the other hand, the number of degrees of freedom in the coarse
grained model is substantially reduced. Each base-sugar-phosphate
residue has only $9$ degrees of freedom in our model once rigid
constraints are accounted for, whereas the same residue has
approximately $100$ degrees of freedom in the all-atom model, if
hydrogen atoms are treated as rigidly bonded. Implicit solvent can be
present in both coarse-grained and all-atom scenarios.  The ten-fold
reduction in the number of degrees of freedom allows the
coarse-grained model to explore longer time scale phenomena than would
be practically obtainable with all-atom simulations.

Base-base interactions had to be strengthened to reproduce the
properties of the crystal structure. Since the putative strength of
the base-base interactions resulted from direct all-atom
parameterization, this implies that the functional form of the RE$^2$
potentials may not fully capture the electron correlations governing
stacking interactions. These polarization effects likely induce a
many-body cooperative component to the stacking interaction for
coarse-grained bases, resulting in a much stiffer potential surface
for local fluctuations around the native structure. Similarly,
many-body interactions between coarse-grained residues in proteins
were necessary to effectively capture protein folding rates and
mechanisms~\citep{EjtehadiMR04}.

We found that the physicochemical based model showed energetic
bias towards towards a right handed form of ds-DNA helix over a
left handed form. The more stable right handed form had longer
persistence length. While the dihedral potentials (as
determined from the equilibrium sampling of a short piece of
ss-DNA) yield potentials that break chiral symmetry towards the
right-handed helix, their role in determining chirality is
sufficiently coupled to other interactions in the system that
strengthening the dihedral potential alone does not enhance
chirality. The chiral energetic bias was largest for the
all-atom parameterized values of the dihedral strength, i.e.
while decreasing the overall strength of the dihedral
potentials diminished chiral preference in the model,
increasing dihedral strength also did not enhance chirality,
but instead diminished it. This effect is likely due to an
interplay between stacking and dihedral interactions. It
appears that like twist, chirality is induced not so much by
the direct inherent energetic preferences of potentials, but by
an indirect minimization of frustration induced by the forced
compression of the system due to base-base stacking
interactions. 

That is, preference for a right-handed helix over the left-handed
helix for ds-DNA in our model arises predominantly from the stacking
of chiral constituents: each phosphate-sugar-base-phosphate
constituent is a non-planar object that cannot be transformed by
rotations and translations into its mirror image. Chirality is
enforced by asymmetries in the equilibrium values of the bond and
angle potentials which determine the minimum energy structure of these
molecular constituents.  Specifically, the molecular building block
P$_{3'}$-S-Base-P$_{5'}$ has sugar residue at the chiral center, and
there is no center of inversion or mirror plane. The subsequent
building block is constrained to have its P$_{3'}$ residue at the
position of the previous block's P$_{5'}$ residue, and is coupled to
the previous building block through base-stacking interactions.  In
this way, the handedness of DNA is induced by coupling these chiral
objects together through stacking interactions, analogous to the
mechanism behind the right-handed preference of $\alpha$-helices due
to hydrogen-bond-mediated coupling between chiral L-form amino
acids~\citep{BrandenTooze91}. Phospholipid-nucleoside conjugates have also
been observed to exhibit spontaneous right-handed helix formation,
with no helical preference present for the conjugate with nucleic
acid bases removed~\citep{YanagawaH89}. This again reinforces the idea 
that helicity can be induced by coupling chiral constituents together through an 
achiral force. Such spontaneous formation of handed
helical structures from chiral ingredients is also reminiscent of the
mechanism by which chiral nematogens, interacting through simple
Lennard-Jones-like potentials, form a cholesteric (twisted nematic)
phase in a liquid crystal~\citep{DeGennesPG93}. Exploration of the
origins of DNA handedness including C-G sequence preference and nucleotide
pairing as in Z-DNA, as well as more refined structural studies of the
connection between atomistic and coarse grained models in the context
of chirality, are topics for future work. 

\section*{ACKNOWLEDGEMENTS}
A. M-A. acknowledges support from a Natural Sciences and
Engineering Research Council of Canada (NSERC) graduate
fellowship, J. R. acknowledges support from NSERC, and S. S. P.
acknowledges support from the A. P. Sloan Foundation and NSERC.
We thank Christopher Yearwood and Andre Marziali for
contributions in the early stages of this project, and Erik Abrahamsson
for helpful discussions. 

\renewcommand{\theequation}{A.\arabic{equation}}
\renewcommand{\thesection}{A\Roman{section}}
\renewcommand{\thefigure}{A\arabic{figure}}
\renewcommand{\thetable}{A\arabic{table}}
\setcounter{section}{0}

\section{APPENDIX} \label{secappendix}

\subsection{The RE$^2$ potential between ellipsoids}
\label{sectre2}

The RE$^2$ potential, developed by Babadi \textit{et
  al.}~\citep{BabadiM06}, is a generalization of the ubiquitous
Lennard-Jones interaction to ellipsoidal particles. The extra
rotational degrees of freedom accounted for by the RE$^2$ potential
come with a greater computational burden: the RE$^2$ potential needs
five times the number of input parameters per interaction pair
(including the cutoff distance) as does the Lennard-Jones potential.

The first six of these parameters come from the radii of the two
ellipsoids, expressed as the following shape tensor:
\begin{eqnarray}
{\bf {S}_i} = \mbox{diag}(\sigma^{(i)}_{x},\sigma^{(i)}_{y},\sigma^{(i)}_{z})
\end{eqnarray}

The anisotropic well depths are expressed as the relative potential
well depth tensor. Its entries are dimensionless and are inversely
proportional to the well depths in their respective directions.
\begin{eqnarray}
{\bf E}_i = \mbox{diag}(\epsilon^{(i)}_{x},\epsilon^{(i)}_{y},\epsilon^{(i)}_{z})
\end{eqnarray}

Another input parameter with a dimension of distance is the atomic
interaction radius, $\sigma_c$, which characterizes the distance scale
for interactions between the atomic constituents to be
coarse-grained. The energy scale is given by the Hamaker constant,
$A_{ij}$. Finally, there is the necessary interaction cutoff distance,
$R_{\small \mbox{cut}}$, as the computation time to compute all
O(N$^2$) RE$^2$ interactions is too prohibitive, and is unnecessary
because of how quickly the potential decays with separation.

The RE$^2$ potential between two ellipsoids, labelled as $i = 1, 2$,
can be conveniently written in terms of attractive and repulsive
components. In the expressions below, the tensor $\mathbf{A}_i$ is the
rotation matrix from the lab frame to the rotated principal axis frame
of particle $i$.
\begin{eqnarray}
V_A^{RE^2}(\mathbf{A}_1, \mathbf{A}_2, \rl{\mathbf{r}}) &=& -\frac{A_{12}}{36}\Big(1+
3\rl{\eta}\rl{\chi}\frac{\sigma_c}{\rl{h}}\Big) \label{eqre2a}  \nn
& &\times
\prod_{i=1}^2\prod_{e=x,y,z}\Bigg(\frac{\sigma^{(i)}_e}{\sigma^{(i)}_e+h_{12}/2}\Bigg)
\label{eqre2b}\\
V_R^{RE^2}(\mathbf{A}_1, \mathbf{A}_2, \rl{\mathbf{r}})&=&\frac{A_{12}}{2025}
\Big(\frac{\sigma_c}{\rl{h}}\Big)^6\Big(1+
\frac{45}{56}\rl{\eta}\rl{\chi}\frac{\sigma_c}{\rl{h}}\Big)\nn
& &\times\prod_{i=1}^2 \prod_{e=x,y,z}
\Bigg(\frac{\sigma^{(i)}_e}{\sigma^{(i)}_e+h_{12}/60^{\frac{1}{3}}}\Bigg)
\end{eqnarray}
The quantities $\eta_{1 2}$, $\chi_{12}$ and $h_{12}$ in
Eqs.~(\ref{eqre2a}-\ref{eqre2b}) are defined by
\begin{eqnarray}
 \eta_{1 2} ({\bf A}_1, {\bf A}_2) &=&
   \frac{ \det[{\bf S}_1]/\sigma_1^{2}+
          \det[{\bf S}_2]/\sigma_2^{2}
        }
        {\left[\det[{\bf H}_{12}]/(\sigma_1+\sigma_2)\right]^{1/2}}
\nonumber
\\
{\bf H}_{1 2}({\bf A}_1,{\bf A}_2, \hat{\bf r}) &=&
\frac1{\sigma_1}{\bf A}_1^T {\bf S}_1^2  {\bf A}_1 +
\frac1{\sigma_2}{\bf A}_2^T {\bf S}_2^2  {\bf A}_2 \nonumber \\
\sigma_i({\bf A}_i, \hat{\bf r}_{1 2}) &=&
\left(\hat{\bf r}_{1 2}^T\ \ {\bf A}_i^T {\bf S}_{i}^{-2}{\bf A}_i
 \hat{\bf r}_{1 2}\right)^{-1/2}
\nonumber
\end{eqnarray}
\begin{eqnarray}
 \chi_{12} &=& 2 \rhat_{12}^T \textbf{B}^{-1}\rhat_{12}\nonumber \\
 \rhat_{12} &=& \textbf{r}_{12}/|\textbf{r}_{12}|	\nonumber \\
{\bf B}&=& {\bf A}_1^T {\bf E_1} {\bf A}_1 + {\bf A}_2^T {\bf E_2} {\bf A}_2
\nonumber
\end{eqnarray}

\begin{eqnarray}
 h_{12} 		&=& |\rhat_{12}| - \sigma_{12}\nonumber \\
\sigma_{12} 	&=& \left[\frac{1}{2} \rhat_{12}^T {\bf G}^{-1} \rhat_{12} \right]^{-1/2} \nonumber \\
{\bf G} 	&=& {\bf A}_1^T {\bf S}_1^2 {\bf A}_1 + {\bf A}_2^T {\bf S}_2^2
{\bf A}_2
\nonumber
\end{eqnarray}

In LAMMPS, an
RE$^2$-like interaction between an ellipsoid and a sphere may be
specified by making the radii parameters of the sphere
zero. The radius of the sphere is given by $\sigma_c$. In this
case, the RE$^2$ interaction between the objects is calculated in the
limit that ${\bf S}_2 \rightarrow {\bf 0}$ and $A_{12} \rightarrow
\infty$ at a rate of $A_{12} \sim 1/\det({\bf S}_2)$. The relevant
energy parameter for this interaction we will call $\tilde{A}_{12}
\equiv A_{12}/\rho \sigma_c^3$, where $\rho$ is the number density of
the sphere. That is, $\frac{4}{3}\pi\det(\mathbf{S}_2)\rho = 1$. The
potential may then be straightforwardly evaluated by substituting
$\mathbf{A}_2 = \mathbf{I}$ and $\mathbf{S}_2 = \mathbf{0}$ into the
RE$^2$ potential:

\begin{eqnarray}
\tilde{V}_A(\mathbf{A}_1, \mathbf{I}, \rl{\mathbf{r}}) &=& -\frac{\tilde{A}_{12}}{36}\frac{4 \sigma_c^3}{3 \pi} \frac{8}{h_{12}}\Big(1+
3\rl{\chi}\frac{\sigma_c}{\rl{h}}\Big) \label{eqnsphereasphereA}
 \prod_{e=x,y,z}\Bigg(\frac{\sigma^{(1)}_e}{\sigma^{(1)}_e+h_{12}/2}\Bigg)\\
\tilde{V}_R(\mathbf{A}_1, \mathbf{I}, \rl{\mathbf{r}})&=&\frac{\tilde{A}_{12}}{2025} \frac{4 \sigma_c^3}{3 \pi} \frac{60}{h_{12}}
\Big(\frac{\sigma_c}{\rl{h}}\Big)^6\Big(1+
\frac{45}{56}\rl{\chi}\frac{\sigma_c}{\rl{h}}\Big)
 \prod_{e=x,y,z}
\Bigg(\frac{\sigma^{(1)}_e}{\sigma^{(1)}_e+h_{12}/60^{\frac{1}{3}}}\Bigg) \label{eqnsphereasphereR}
\end{eqnarray}

A more complete discussion of the RE$^2$ potential including its
advantages over the alternative biaxial Gay-Berne
potential~\citep{BerardiR98} is discussed in the
literature~\citep{BabadiM06}.

\subsection{Energetic and Hydrodynamic comparison.}
\label{sectEHcompare}

For our Langevin simulations, we directly use the friction eigenvalues
extracted from all-atom simulations.  However, it is instructive to
compare the effective size of the bases both energetically and
hydrodynamically.

The shape-dependent friction coefficients of ellipsoidal bodies
can be obtained from the low Reynolds number solution to the
Navier-Stokes hydrodynamics equations \citep{lamb1932h}, and
can be expressed in terms of the three radii $r_i$ in the
principal axes: \alpheqn
\begin{eqnarray}
\zeta_{i}^{\left(t\right)}&=&\frac{16\pi\eta}{S+G_i}
\label{eqzetatrans} \\
\zeta_{i}^{\left(r\right)}&=&\frac{16\pi\eta(r_j^2+r_k^2)}{3(G_j+G_k)},
\label{eqzetarot}
\end{eqnarray}
\reseteqn
with the elliptical integrals
$$
S=\int_0^\infty d\lambda
\left[(r_1^2+\lambda)(r_2^2+\lambda)(r_3^2+\lambda)\right]^{-1/2}
$$
and \\
$$
G_i = r_i^2\int_0^\infty d\lambda (r_i^2+\lambda)^{-1}
\left[(r_1^2+\lambda)(r_2^2+\lambda)(r_3^2+\lambda)\right]^{-1/2} \: .
$$

To verify that the continuum hydrodynamic expressions for the friction
tensor faithfully reproduce the diffusive motion of a base, we
extracted diffusion coefficients from all atom simulations using the
CHARMM potential. The all atom simulations were performed with a
single base solvated in a (20 \AA)$^3$ box of water in a 0.2 mol/L
neutral KCl solution. The simulations were performed for $10$ ns in
the NPT ensemble. Four such simulations were performed, one for each
type of base. Temperature was maintained at $310$K by Langevin
dynamics and pressure was maintained at $1$ atm.

Hydrodynamic radii are found by best fit of Eqs.~(\ref{eqzetatrans}-b)
to the observed effective friction coefficients for both rotation and
translation in all-atom simulations.  Tabulated values of friction
coefficients and effective hydrodynamic radii are shown in
Table~\ref*{S1-tabfrict}. These hydrodynamic radii may then be
compared with energetic radii (for our Langevin simulations, we
directly use the friction eigenvalues extracted from all-atom
simulations rather than the parameters resulting from the best fit to
the hydrodynamics of an ellipsoid).

\begin{figure}[!h]
\centering
\includegraphics*[width=8cm]{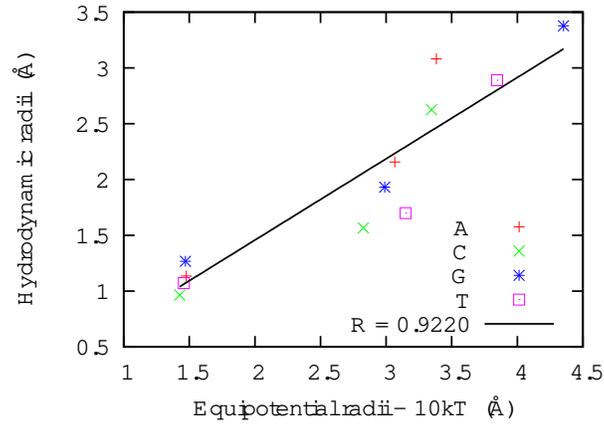}
\caption[]{Hydrodynamic vs. energetic radii for effective
  ellipsoids. Hydrodynamic radii are found by optimizing the measured
  friction coefficients extracted from all-atom simulations of
  isolated bases and fitting to Eqs.~(\ref{eqzetatrans}-b) for the
  rotational and translational friction of an ellipsoid.  The values
  are strongly correlated, however hydrodynamically-derived radii tend
  to be smaller than energetically-derived radii. For the thinnest
  axis, the assumption of continuum hydrodynamics is expected to be
  least accurate, and relative modifications due to hydration effects
  are expected to be
  largest~\citep{SchneiderB95,ElcockAH95,FeigM99,delaTorreJ00,HalleB03}.}
\label{figHEradii}
\end{figure}

A plot of hydrodynamic radii {\it vs.} energetic radii is shown in
Fig.~\ref{figHEradii}. The two measures compare well, however the
hydrodynamic size tends to be smaller than the energetic size. Either
the parameterization scheme we employed resulted in RE$^2$ potentials
that overestimated the range of repulsive forces, or the smaller
hydrodynamic values may be the result of the breakdown of the no-slip
condition at the interface between bases and water, a reasonable
scenario given that the size scale of the bases ($\sim$ \AA) tests the
limits of the macroscopic assumptions in continuum hydrodynamics.

\clearpage
\subsection{Figures}

\begin{figure}[!h]
\centering
	\includegraphics*[width=8cm]{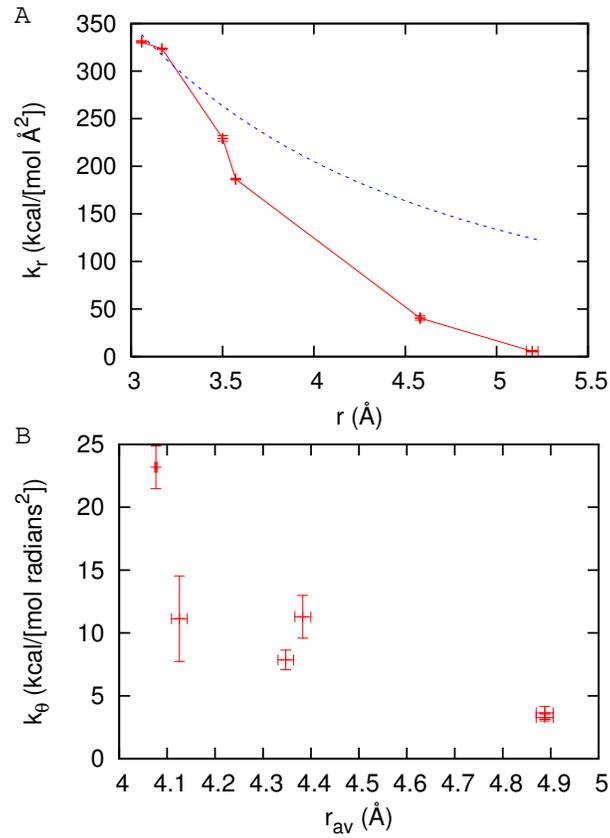}
\caption[]{(A) Stiffness coefficient $k_r$ for bond potentials as a
  function of the mean distance between the participating residues,
  along with the phonon dispersion relation for a 1-D chain of
  harmonic oscillators~\citep{AshcroftNW76}, $k_{eff}(\lambda) = k_o \sin^2
  (\pi a/\lambda)$, with wavelength $\lambda$ here taken to be the distance
  $r$. The correlation coefficient is $r=0.995$ and chance probability
  $p\sim4\times 10^{-5}$. (B) Stiffness coefficient $k_\theta$ for
  angle potentials as a function of the mean length of the two bonds
  participating in the angle.}
      \label{figkvsr}
      \label{figkvsrtheta}
\end{figure}

\begin{figure}[!h]
\begin{center}
\includegraphics[width=8cm]{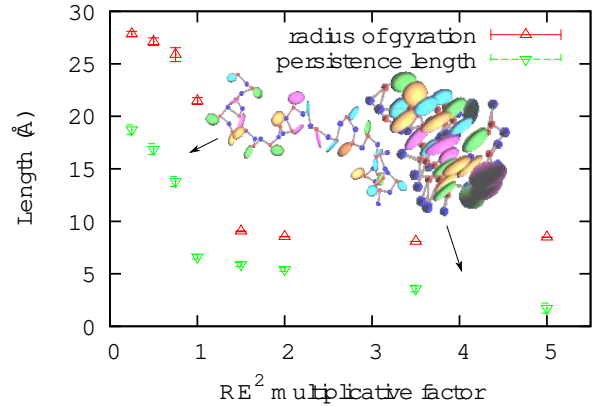}
\end{center}
\caption[]{Persistence length $\lp$ and radius of gyration $R_g$ vs. a
  multiplicative factor weighting the RE$^2$ energies for a ssDNA
  sequence $HET_{SS,25}$ at ionic concentration of 0.04 mol/L,
  temperature $0.42\epsilon$ simulated for 240 ns.  Note that the
  parameters, resulting from fitting the RE$^2$ potential to all-atom
  potentials, result in DNA whose persistence length and collapse is
  highly sensitive to external conditions. Such a sensitivity could be
  exploited {\it in vivo} by either varying the ionic environment or
  through histone phosphorylation, acetylation, or methylation, for
  the purposes of gene regulation.  Insets show representative
  snapshots taken from the simulations~\citep{abrahamsson2009biovec}.
}
\label{figsingleRGvRE2}
\end{figure}

\begin{figure}[!h]
\begin{center}
\includegraphics[width=8cm]{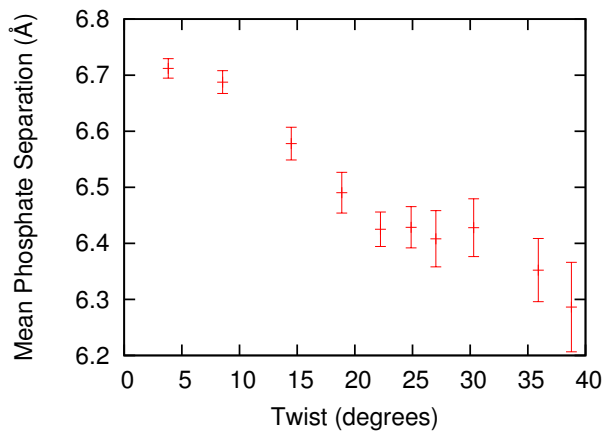}
\end{center}
\caption[]{ Mean distance between phosphate groups as a function of
  helical twist. Data taken from the same simulations as in
  Fig.~\ref{figstackpitchre2} }
\label{figmeanPhosphateTwist}
\end{figure}

\end{document}